\documentclass[twocolumn,times]{aastex62}

\usepackage{textcomp}
\usepackage{natbib}
\usepackage{epsfig,amsmath,amssymb}
\usepackage{graphicx}
\usepackage{tabularx}
\usepackage{array}

\def\CII{[C\,\textsc{ii}]}
\def\logNHI[#1]{$\log(N_{\rm H\scriptscriptstyle{I}}/{ \rm cm^{-2}})$ = #1}
\def\kms{km~s$^{-1}$}
\def\LTIR{$L_{\rm TIR}$}
\def\LCII{$L_{\rm [C{\scriptscriptstyle II}]}$}
\def\sfrunit{$M_\odot$~yr$^{-1}$}

\shorttitle{Resolved \CII\ Emission from $z > 6$ Quasar -- Galaxy Pairs} 
\shortauthors{Neeleman et al.}
 
\begin{document}
\title{\Large Resolved \CII\ Emission from $z > 6$ Quasar Host -- Companion Galaxy Pairs}

\correspondingauthor{Marcel Neeleman}
\email{neeleman@mpia.de}

\author[0000-0002-9838-8191]{Marcel Neeleman}
\affiliation{Max-Planck-Institut f\"{u}r Astronomie, K\"{o}nigstuhl 17, D-69117, Heidelberg, Germany}
\author[0000-0002-2931-7824]{Eduardo Ba\~{n}ados}
\affiliation{Max-Planck-Institut f\"{u}r Astronomie, K\"{o}nigstuhl 17, D-69117, Heidelberg, Germany}
\author[0000-0003-4793-7880]{Fabian Walter}
\affiliation{Max-Planck-Institut f\"{u}r Astronomie, K\"{o}nigstuhl 17, D-69117, Heidelberg, Germany}
\affiliation{National Radio Astronomy Observatory, Socorro, NM 87801, USA}
\author[0000-0002-2662-8803]{Roberto Decarli}
\affiliation{INAF -- Osservatorio di Astrofisica e Scienza dello Spazio di Bologna, via Gobetti 93/3, I-40129, Bologna, Italy}
\author[0000-0001-9024-8322]{Bram P. Venemans}
\affiliation{Max-Planck-Institut f\"{u}r Astronomie, K\"{o}nigstuhl 17, D-69117, Heidelberg, Germany}
\author[0000-0001-6647-3861]{Chris L. Carilli}
\affiliation{National Radio Astronomy Observatory, Socorro, NM 87801, USA}
\affiliation{Cavendish Astrophysics Group, University of Cambridge, Cambridge CB3 0HE, UK}
\author[0000-0003-3310-0131]{Xiaohui Fan}
\affiliation{Steward Observatory, University of Arizona, 933 North Cherry Avenue, Tucson, AZ 85721, USA}
\author[0000-0002-6822-2254]{Emanuele P. Farina}
\affiliation{Max-Planck-Institut f\"{u}r Astronomie, K\"{o}nigstuhl 17, D-69117, Heidelberg, Germany}
\affiliation{Max-Planck-Institut f\"{u}r Astrophysik, Karl-Schwarzschild-Strasse 1, D-85748, Garching, Germany}
\author[0000-0002-5941-5214]{Chiara Mazzucchelli}
\affiliation{European Southern Observatory, Alonso de C\'{o}rdova 3107, Vitacura, Regi\'{o}n Metropolitana, Chile}
\author{Mladen Novak}
\affiliation{Max-Planck-Institut f\"{u}r Astronomie, K\"{o}nigstuhl 17, D-69117, Heidelberg, Germany}
\author[0000-0001-9585-1462]{Dominik A. Riechers}
\affiliation{Cornell University, Space Sciences Building, Ithaca, NY 14853, USA}
\affiliation{Max-Planck-Institut f\"{u}r Astronomie, K\"{o}nigstuhl 17, D-69117, Heidelberg, Germany}
\author[0000-0003-4996-9069]{Hans-Walter Rix}
\affiliation{Max-Planck-Institut f\"{u}r Astronomie, K\"{o}nigstuhl 17, D-69117, Heidelberg, Germany}
\author[0000-0003-4956-5742]{Ran Wang}
\affiliation{Kavli Institute of Astronomy and Astrophysics at Peking University, No.5 Yiheyuan Road, Haidan District, Beijing, 100871, China}

\begin{abstract}
We report on $\approx$$0.35\arcsec$($\approx$$2$kpc) resolution observations of the \CII\ and dust continuum emission from five $z>6$ quasar host -- companion galaxy pairs obtained with the Atacama Large Millimeter/submillimeter Array. The \CII\ emission is resolved in all galaxies, with physical extents of $3.2 - 5.4$~kpc. The dust continuum is on-average 40\% more compact, which results in larger \CII\ deficits in the center of the galaxies. However, the measured \CII\ deficits are fully consistent with those found at lower redshifts. Four of the galaxies show \CII\ velocity fields that are consistent with ordered rotation, while the remaining six galaxies show no clear velocity gradient. All galaxies have high ($\sim$$80 - 200$ \kms) velocity dispersions, consistent with the interpretation that the interstellar medium (ISM) of these high redshift galaxies is turbulent. By fitting the galaxies with kinematic models, we estimate the dynamical mass of these systems, which range between ($0.3$ $-$ $>$$5.4) \times 10^{10}~M_\odot$. For the three closest separation galaxy pairs, we observe dust and \CII\ emission from gas in between and surrounding the galaxies, which is an indication that tidal interactions are disturbing the gas in these systems. Although gas exchange in these tidal interactions could power luminous quasars, the existence of quasars in host galaxies without nearby companions suggests that tidal interactions are not the only viable method for fueling their active centers. These observations corroborate the assertion that accreting supermassive black holes do not substantially contribute to the \CII\ and dust continuum emission of the quasar host galaxies, and showcase the diverse ISM properties of galaxies when the universe was less than one billion years old.
\end{abstract}


\section{Introduction}
One of the hallmarks of $\Lambda$ cold dark matter ($\Lambda$-CDM) cosmology is the hierarchical formation of galaxies, whereby galaxies acquire their mass through a sequence of mergers and mass inflows \citep[e.g.,][]{Keres2005,Dekel2009}. In this paradigm, mergers, in particular major mergers, are often attributed to be the dominant process that creates massive, quiescent galaxies \citep{Vanderwel2009}. Recent observations of massive, quiescent galaxies at high redshifts \citep[$z \gtrsim 3$; e.g.,][]{Straatman2014,Glazebrook2017} therefore necessitates that a significant fraction of major mergers must have occurred at even higher redshifts ($z > 6$). 

Such mergers are likely to occur in large galaxy overdensities, as the elevated density increases the likelihood of two galaxies interacting. This is also where one would expect to find the most luminous quasars, as luminous accreting supermassive black holes are hosted by massive galaxies \citep[e.g.,][]{Targett2012}, and massive galaxies trace galaxy overdensities. It is therefore not surprising that, at low redshift, luminous quasars occur more frequently in merging galaxies \citep{Hong2015}, especially since mergers could provide the fuel that triggers the quasar-phase of the supermassive black hole. 

At $z >6$, dependencies between luminous quasars and their environs are still debated. To study the environment of high redshift quasars, predominantly optical and near-infrared imaging has been done \citep[e.g.,][]{Kim2009,Banados2013,McGreer2014,Goto2017,Mazzucchelli2017,Ota2018,Mazzucchelli2019}. Results from these studies are inconclusive, with some studies finding evidence for galaxy overdensities \citep[e.g.,][]{Garcia-Vergara2017}, while others find no overdensities or even underdensities \citep[e.g.,][]{Uchiyama2018} around luminous quasars. This discrepancy is attributed to both the redshift uncertainty of the quasar as well as survey depths \citep[see the discussion in][]{Champagne2018}. Together with the intrinsic brightness of the quasar in the optical/near-infrared, studying the immediate surroundings of the quasar to detect, let alone spectroscopically confirm, any putative nearby galaxies remains extremely challenging, with only limited success \citep[e.g.,][]{Farina2017, Mazzucchelli2019}. 

Fortunately, complementary to optical and near-infrared imaging, we can study the quasars in the millimeter regime. In particular, the fine-structure line of singly ionized carbon at 157.7$\mu$m (\CII) has been used extensively to study the interstellar medium (ISM) of high-redshift objects \citep[e.g.,][]{Carilli2013}. In \citet{Decarli2018}, the \CII\ emission line was used to study a sample of 27 $z > 6$ quasars using the Atacama Large Millimeter/submillimeter Array (ALMA). This program, aimed at characterizing the \CII\ emission line of high redshift quasars using short ($\approx$8 min on-source), $\sim$1$\arcsec$-resolution ALMA observations, revealed that at least 4 out of 27 quasar fields contained a bright \CII-emitting companion \citep{Decarli2017}. This fraction is well-above what is expected from number counts of \CII\ emitters, but is consistent with expectations, if galaxy clustering is taken into account. 

Similar companion galaxies have also been found around $z \sim 5$ redshift quasars \citep{Trakhtenbrot2017}, and more generally around other high redshift, rest-frame far-infrared bright sources \citep[e.g.,][]{Omont1996,Ivison2012,Oteo2016,Riechers2017,Marrone2018}. Moreover, deeper, higher resolution observations will likely reveal even more companion galaxies. However, these brightest \CII\ companion galaxies near $z > 6$ quasars are remarkable, as their \CII\ characteristics are comparable to those of the quasar host galaxies. This results in similar physical properties for both the quasar host and these companion galaxies. Their close separation ($<$65~kpc) further indicates that these galaxy pairs are interacting or will interact in the future, suggesting that they are the precursors of $z \gtrsim 3$, massive, quiescent galaxies.

Facilitated by the strength of the \CII\ line, we have obtained $\approx$$0.35\arcsec$ resolution \CII\ observations of five, \CII\-bright, $z >6$ quasar host -- companion galaxy pairs. These higher resolution observations ($\sim$10$\times$ smaller beam area) resolve the \CII\ emission at $\approx$2~kpc scales at the redshift of the galaxies, yielding detailed information on the structure and kinematics of the ISM in these galaxies. In addition, these observations allow us to study the gas interactions between the quasar host and companion galaxy for galaxy pairs at different angular separation, and allow us to determine if the accreting supermassive black holes in the quasar host galaxies alter the observed far-infrared properties of their host galaxies. Throughout this paper we assume a standard, flat, $\Lambda$-CDM concordance cosmology with $\Omega_\lambda = 0.7$ and $H_0 = 70~\text{km}~\text{s}^{-1}~\text{Mpc}^{-1}$.

\section{Observations and Data Reduction}
The five quasar fields discussed in this paper were observed with ALMA in Cycle 4 and 5 (proposal IDs: 2016.1.00544.S, 2017.1.01301.S). Four of these companion galaxies were previously known \citep{Decarli2017,Willott2017}. The remaining system (J1306$+$0356) was previously identified as a spatially-resolved source \citep{Decarli2018}, but the higher resolution data clearly separates the \CII\ emission into two distinct sources. We note that we selected these five quasar-companion galaxy pairs from the larger quasar sample solely by the \CII\ emission strength of both the quasar host and the companion, in order to efficiently study these pairs in reasonable observing times. It is very likely that the other quasars have companion galaxies, which are either much fainter or closer ($\lesssim 1 \arcsec$) to the quasar.

\begin{table*}[!t]
\centering
\caption{Description of ALMA Observations
\label{tab:obs}}
\newcolumntype{C}{>{\centering\arraybackslash}X}%
\begin{tabularx}{\textwidth}{CCCCCC}
\hline
\hline
& J0842$+$1218 & PJ167$-$13 & J1306$+$0356 & PJ231$-$20 & J2100$-$1715\\
\hline
R. A. (J2000) & 08:42:29.20 & 11:10:33.98  & 13:06:08.26 &  15:26:37.84 & 21:00:54.62\\
Dec. (J2000) & $+$12:18:52.5 & $-$13:29:45.6 & $+$03:56:26.3 & $-$20:50:00.7 & $-$17:15:22.5\\
Observed frequency (GHz)$^{\rm a}$ & 268.619 & 252.877 & 270.050 & 250.539 & 269.001\\
Total on-source time (min) & 54.4 & 43.3 & 28 .2 & 43.8 & 25.7\\
Cont. resolution ($\arcsec \times \arcsec$)$^{\rm b}$ & $(0.38 \times 0.30)$ & $(0.42 \times 0.33)$ & $(0.40 \times 0.36)$ & $(0.30 \times 0.25)$ & $(0.36 \times 0.29)$\\
Continuum RMS ($\mu$Jy beam$^{-1}$) & 16.3 & 11.5 & 23.4 & 12.3 & 21.1\\
Cube resolution ($\arcsec \times \arcsec$)$^{\rm b}$ & ($0.36 \times 0.28$) & ($0.43 \times 0.32$) & ($0.40 \times 0.34$) & ($0.29 \times 0.23$) & ($0.35 \times 0.28$)\\
Channel width (km~s$^{-1}$) & 34.9 & 37.0 & 34.7 & 37.4 & 34.9\\
RMS per channel (mJy beam$^{-1}$) & 0.23 & 0.16 & 0.33 & 0.17 & 0.33\\
\hline
\multicolumn{6}{l}{$^{\rm a}$ Central frequency of the spectral window containing the \CII\ line.}\\
\multicolumn{6}{l}{$^{\rm b}$ Full width at half maximum of the synthesized beam of the image.}\\
\end{tabularx}
\end{table*}

For the frequency setup, one of the spectral windows was tuned to contain both the \CII\ emission of the quasar host and the companion galaxy, except for J2100$-$1715 in which two adjoining spectral windows were centered on the \CII\ line. The remaining bands were set up to detect far-infrared (FIR) continuum emission from the field. The observations were performed in array configurations with median baselines of $\approx$210~m for J0842$+$1218 and PJ167$-$13, and median baselines of $\approx$340~m for the remaining three fields. Total exposure times and array setups are given in Table \ref{tab:obs}.

The observations were initially processed using the ALMA pipeline, which is part of the Common Astronomy Software Application package \citep[CASA;][]{McMullin2007}. The calibrated data were then combined with the previous short ($\approx$8 min) compact array observations (PID: 2015.1.01115.S), although we note that this only marginally changed the sensitivity of the observations. Minor additional flagging of several atmospheric lines was done on the combined data set. For the quasar field toward PJ231$-$20, the continuum flux of the quasar was bright enough to perform several rounds of phase-only self-calibration. All this was done using standard routines available in CASA. The continuum images were created using natural weighting with \texttt{tclean} from those channels that showed no significant line emission. To create the spectral line cubes, we subtracted the continuum in the \textit{uv}-plane using the \texttt{uvcontsub} routine. The spectral window(s) that contains both the \CII\ emission from the quasar host as well as the companion galaxy was then spectrally averaged over 16 channels ($\approx$31~MHz or $\approx$35 \kms), and imaged using natural weighting with \texttt{tclean}. Resolution and root-mean-square (RMS) sensitivities for the resulting images are tabulated in Table \ref{tab:obs}. 

We note that we also imaged the data with a robust weighting scheme with a Briggs parameter of 0.5. However, for the final analysis, we opted for the increased signal-to-noise of the natural weighting scheme compared to slightly better resolution of the robust weighting scheme. This choice does not affect the analysis presented in this paper.

\begin{deluxetable*}{lcccccccc}
\tablecaption{Observed Far-Infrared Properties of the Quasar Host and Companion Galaxies
\label{tab:obsfirprop}}
\tablehead{
\colhead{Name\tablenotemark{a}} & 
\colhead{R.A.} &
\colhead{Dec.} &
\colhead{$z_{\rm [C\scriptscriptstyle{II}]}$\tablenotemark{b}} &
\colhead{$F_{\rm [C\scriptscriptstyle{II}]}$\tablenotemark{c}} &
\colhead{FWHM$_{\rm [C\scriptscriptstyle{II}]}$\tablenotemark{d}} &
\colhead{$F_{\rm cont}$\tablenotemark{e}} &
\colhead{$A_{\rm cont}$\tablenotemark{f,h}} &
\colhead{$A_{\rm [C\scriptscriptstyle{II}]}$\tablenotemark{g,h}} \\
\colhead{} & 
\colhead{(J2000)} &
\colhead{(J2000)} &
\colhead{} &
\colhead{(Jy~\kms)} &
\colhead{(\kms)} &
\colhead{(mJy)} &
\colhead{($\arcsec \times \arcsec$)} &
\colhead{($\arcsec \times \arcsec$)}
} 
\startdata
J0842$+$1218Q & 08:42:29.439 & $+$12:18:50.50 & $6.0760(2)$ & $1.40 \pm 0.14$ & $390 \pm 40$ & $0.72 \pm 0.06$ & $(0.28 \times 0.16)$ & $(0.57 \times 0.37)$\\ 
J0842$+$1218C & 08:42:28.967 & $+$12:18:54.98 & $6.0656(3)$ & $1.80 \pm 0.24$ & $310 \pm 30$ & $0.23 \pm 0.05$ & $<$$(0.8 \times 0.5)$\tablenotemark{i} & $(0.48 \times 0.30)$\\
J0842$+$1218C2 & 08:42:29.674 & $+$12:18:46.28 & $6.0649(3)$ & $0.41 \pm 0.09$ & $280 \pm 50$ & $<$$0.032$\tablenotemark{j} & --- & $<$$(0.9 \times 0.7)$\\ 
\hline
PJ167$-$13Q & 11:10:33.979 & $-$13:29:45.82 & $6.5154(3)$ & $3.32 \pm 0.14$ & $490 \pm 30$ & $0.71 \pm 0.05$ & $(0.50 \times 0.28)$ & $(0.99 \times 0.57)$\\ 
PJ167$-$13C & 11:10:34.033 & $-$13:29:46.29 & $6.5119(3)$ & $1.24 \pm 0.09$ & $460 \pm 40$ & $0.16 \pm 0.03$ & $(0.31 \times 0.30)$ & $(0.66 \times 0.37)$\\ 
\hline
J1306$+$0356Q & 13:06:08.261 & $+$03:56:26.26 & $6.0328(3)$ & $1.80 \pm 0.15$ & $278 \pm 27$ & $0.92 \pm 0.09$ & $(0.36 \times 0.30)$ & $(0.56 \times 0.42)$\\ 
J1306$+$0356C & 13:06:08.324 & $+$03:56:26.18 & $6.0342(4)$ & $0.86 \pm 0.11$ & $180 \pm 30$ & $0.30 \pm 0.09$ & $(0.33 \times 0.17)$ & $(0.64 \times 0.49)$\\ 
\hline
PJ231$-$20Q & 15:26:37.837 & $-$20:50:00.75 & $6.5867(3)$ & $5.67 \pm 0.29$ & $411 \pm 20$ & $4.03 \pm 0.12$ & $(0.13 \times 0.10)$ & $(0.35 \times 0.30)$\\ 
PJ231$-$20C & 15:26:37.872 & $-$20:50:02.36 & $6.5901(3)$ & $3.12 \pm 0.26$ & $440 \pm 30$ & $1.26 \pm 0.11$ & $(0.38 \times 0.31)$ & $(0.41 \times 0.40)$\\ 
PJ231$-$20C2 & 15:26:37.972 & $-$20:50:02.40 & $6.5906(3)$ & $0.26 \pm 0.07$ & $310 \pm 60$ & $0.19 \pm 0.04$ & $(0.41 \times 0.35)$ & $(0.61 \times 0.47)$\\ 
\hline
J2100$-$1715Q & 21:00:54.702 & $-$17:15:21.96 & $6.0809(3)$ & $2.26 \pm 0.22$ & $360 \pm 50$ & $0.88 \pm 0.09$ & $(0.44 \times 0.25)$ & $(0.58 \times 0.46)$\\ 
J2100$-$1715C & 21:00:55.456 & $-$17:15:22.08 & $6.0814(2)$ & $4.15 \pm 0.52$ & $610 \pm 80$ & $2.25 \pm 0.21$ & $(0.45 \times 0.28)$ & $(0.66 \times 0.31)$
\enddata
\tablecomments{$^{\rm a}$Quasar hosts are appended by a Q after the short name given in \citet{Decarli2018}, whereas companion galaxies are appended by a C. For those fields with a second companion galaxy, the galaxy name is appended with C2. $^{\rm b}$Redshift of the \CII\ line, as determined from a Gaussian fit to the data. Uncertainties in the last digit are given in parentheses. $^{\rm c}$Velocity-integrated \CII\ line flux density. $^{\rm d}$FWHM of the \CII\ line. $^{\rm e}$Continuum flux density. $^{\rm f}$Size of continuum deconvolved from the beam, as determined from a 2D Gaussian fit to the data. $^{\rm g}$Size of \CII\ line deconvolved from the beam, as determined from a 2D Gaussian fit to the data. $^{\rm h}$No uncertainties are given for the size estimates as these estimates could be off by as much as a factor of 2 (see Section \ref{sec:Size}). $^{\rm i}$Source is unresolved. $^{\rm j}$2$\sigma$ upper limit.}
\end{deluxetable*}

\begin{deluxetable*}{lcccccccc}
\tablecaption{Derived Far-Infrared Properties of the Quasar Host and Companion Galaxies
\label{tab:derfirprop}}
\tablehead{
\colhead{Name\tablenotemark{a}} & 
\multicolumn{2}{c}{Impact parameter\tablenotemark{b}} &
\colhead{\LCII\tablenotemark{c}} &
\colhead{\LTIR\tablenotemark{d}} &
\colhead{SFR$_{\rm [C\scriptscriptstyle{II}]}$\tablenotemark{e}} &
\colhead{SFR$_{\rm TIR}$\tablenotemark{f}} &
\colhead{$M_{\rm dyn, obs}$\tablenotemark{g}} &
\colhead{$M_{\rm dyn, mod}$\tablenotemark{h}} \\
\colhead{} & 
\colhead{($\arcsec$)} &
\colhead{(kpc)} &
\colhead{($10^{9} L_\odot$)} &
\colhead{($10^{11} L_\odot$)} &
\colhead{(\sfrunit)} &
\colhead{(\sfrunit)} &
\colhead{($10^{10} M_\odot$)} &
\colhead{($10^{10} M_\odot$)}
} 
\startdata
J0842$+$1218Q & & & $1.35 \pm 0.14$ & $5.1 - 23$ & $71 - 450$ &  $77 - 350$ & $1.5 - 4.8$ & $>$1.3\\
J0842$+$1218C & $8.2$ & $47$ & $1.73 \pm 0.23$ & $1.6 - 7.5$ & $95 - 600$ &  $24 - 110$ & $0.8 - 2.5$ & $>$$2.3$\\
J0842$+$1218C2 & $5.4$ & $31$ & $0.39 \pm 0.09$ & --- & $17 - 100$ &  --- & $<$4 & ---\tablenotemark{i} \\
\hline
PJ167$-$13Q & & & $3.53 \pm 0.15$ & $5.6 - 26$ & $220 - 1400$ &  $84 - 380$ & $4.1 - 13$ & $3.5 \pm 0.4$\\
PJ167$-$13C & $0.92$ & $5.0$ & $1.32 \pm 0.10$ & $1.3 - 5.8$ & $69 - 440$ &  $19 - 86$ & $2.4 - 7.4$ & $>$1.8\\
\hline
J1306$+$0356Q & & & $1.71 \pm 0.14$ & $6.5 - 30$ & $94 - 590$ &  $97 - 440$ & $0.8 - 2.4$ & $>$$0.6$\\
J1306$+$0356C & $0.95$ & $5.4$ & $0.82 \pm 0.10$ & $2.1 - 9.7$ & $39 - 250$ &  $32 - 140$ & $0.4 - 1.1$ & $>$0.3\\
\hline
PJ231$-$20Q & & & $6.13 \pm 0.31$ & $32 - 150$ & $420 - 2700$ &  $480 - 2200$ & $2.0 - 6.2$ & $>$$2.0$\\
PJ231$-$20C & $1.7$ & $9.1$ & $3.37 \pm 0.28$ & $10 - 46$ & $210 - 1300$ &  $150 - 690$ & $2.7 - 8.4$ & $>$$5.4$\\
PJ231$-$20C2 & $2.5$ & $14$ & $0.28 \pm 0.08$ & $1.5 - 7.0$ & $11 - 70$ &  $23 - 100$ & $2.0 - 3.1$ & ---\tablenotemark{i}\\
\hline
J2100$-$1715Q & & & $2.17 \pm 0.21$ & $6.3 - 29$ & $120 - 790$ &  $94 - 430$ & $1.3 - 4.1$ & $>$$1.1$\\
J2100$-$1715C & $10.8$ & $61$ & $3.99 \pm 0.50$ & $16 - 73$ & $260 - 1600$ &  $240 - 1100$ & $4.1 - 14$ & $4.2 \pm 1.0$\\
\enddata
\tablecomments{$^{\rm a}$Name as defined in Table \ref{tab:obsfirprop}. $^{\rm b}$Projected distance from the quasar host to the companion galaxy at the redshift of the quasar. $^{\rm c}$\CII\ luminosity. $^{\rm d}$Total infrared luminosity as calculated from the continuum measurement (Section \ref{sec:IndSor}). $^{\rm e}$SFR determined from the \CII\ luminosity \citep{Delooze2014}. $^{\rm f}$SFR determined from the dust continuum / total infrared luminosity \citep{Kennicutt2012}. $^{\rm g}$Dynamical mass estimate determined from the FWHM of the \CII\ line. $^{\rm h}$Dynamical mass estimate determined from the kinematical modeling of the \CII\ line. $^{\rm i}$No dynamical mass estimate because of limited signal-to-noise.}
\end{deluxetable*}

\begin{figure*}[!t]
\includegraphics[width=\textwidth]{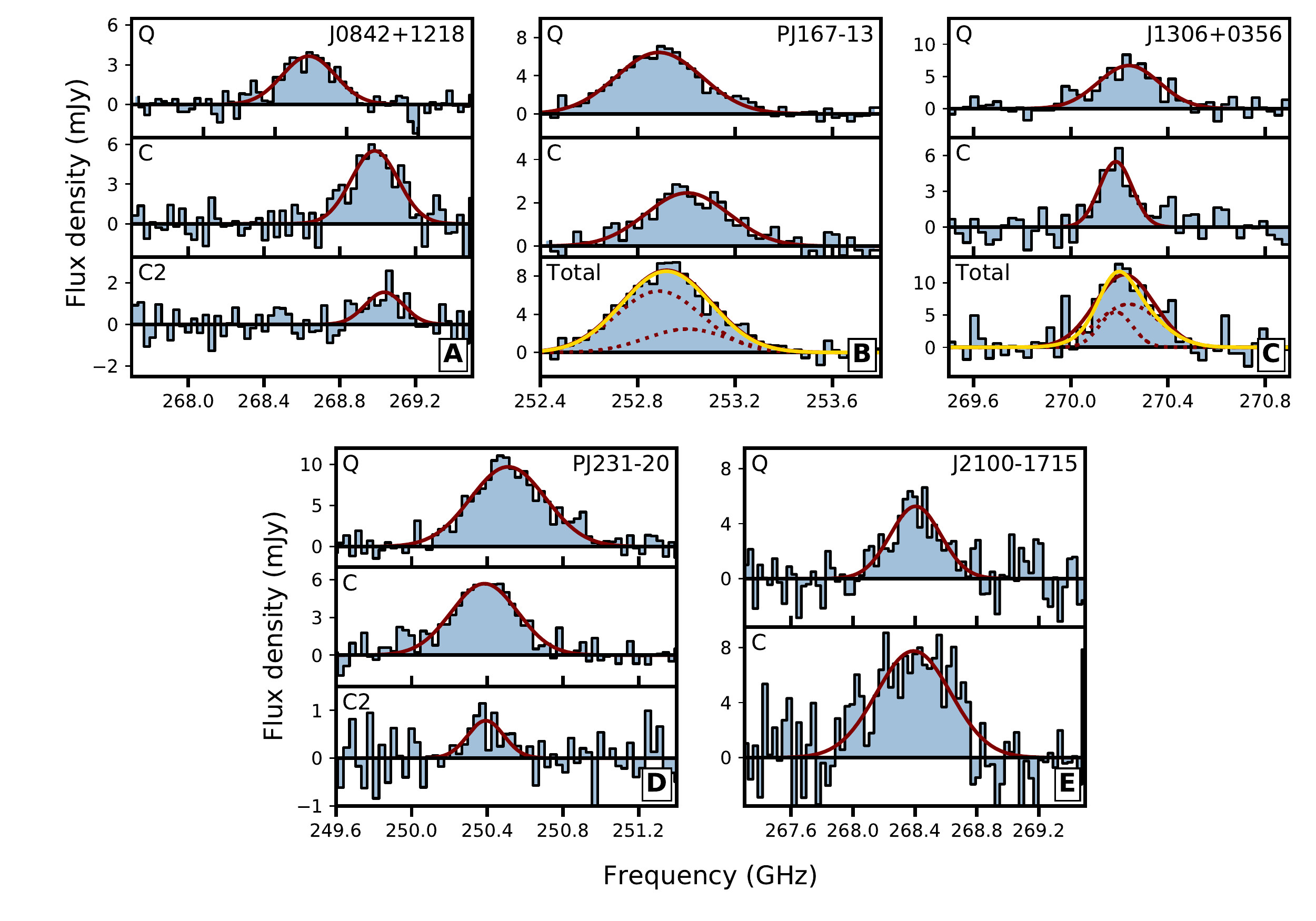}
\caption{Continuum-subtracted spectra of the \CII\ line from the quasar host and companion galaxies extracted from a $1\farcs5$ aperture centered on the emission. The name of the quasar is shown in the top right of each panel, the quasar host (top panel)  and companion galaxy (second panel from top) are marked by a `Q' and `C' , respectively. For those two fields (J0842$+$1218 and PJ231$-$20) that show a second \CII\ companion galaxy, `C2', the bottom panels display the spectrum of these emitters. A Gaussian fit for each emitter is shows as well (solid maroon line). For quasars PJ167$-$13 and J1306$+$0356, previous observations were unable to resolve both components separately, therefore also the total \CII\ flux is shown in the bottom panel. The yellow line in these figures is the sum of the two Gaussian fits from the quasar host and companion galaxy. The excellent fit of these lines to the data obviates the need for any additional emission components.
\label{fig:CIISpec}}
\end{figure*}

\begin{figure*}[!th]
\includegraphics[width=\textwidth]{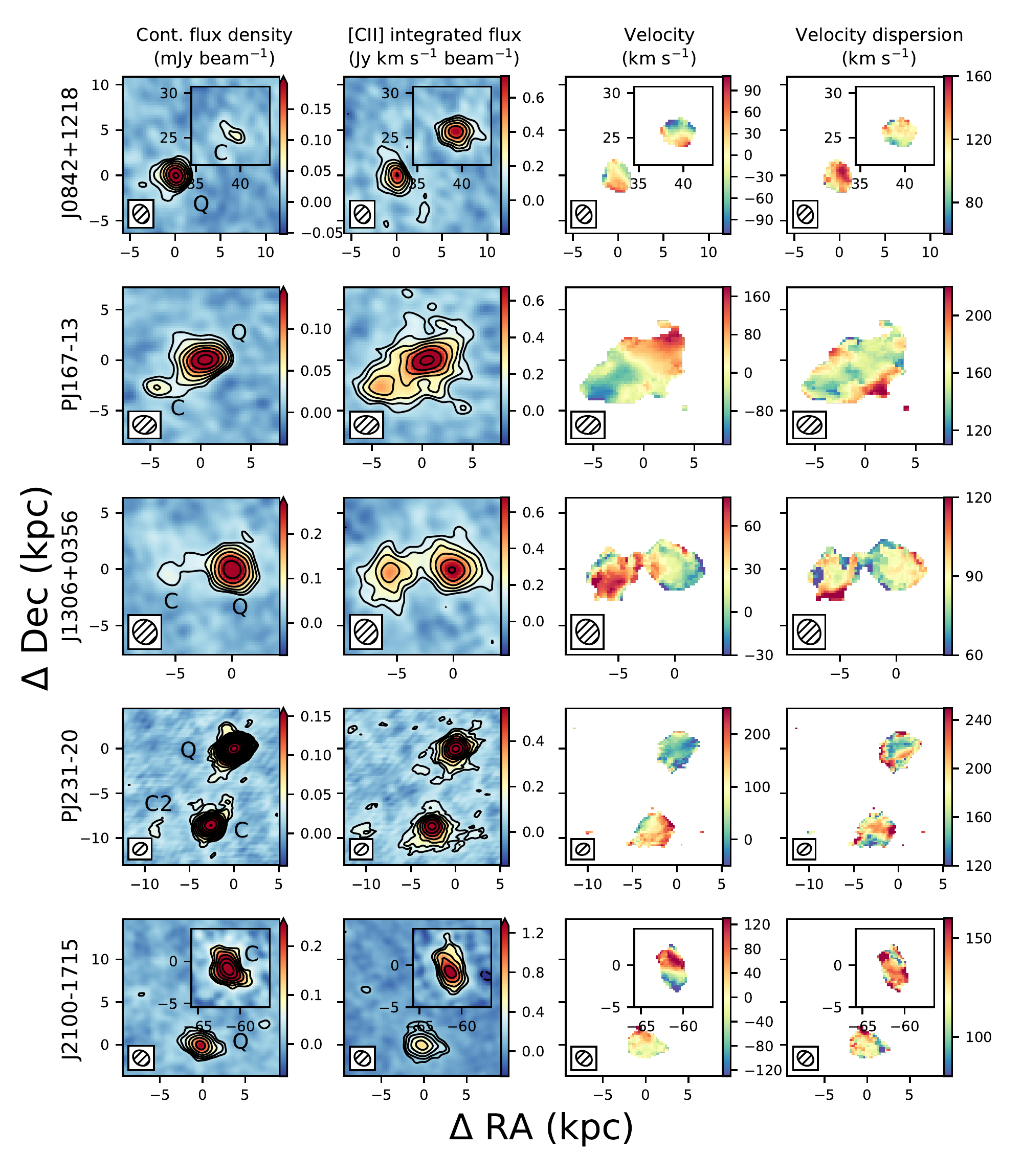}
\caption{ALMA observations of the full sample of quasar host -- companion galaxy pairs. The first column shows the continuum flux density, while the second column shows the integrated \CII\ flux density of the galaxies. The final two columns show the mean velocity field and the velocity dispersion of the \CII\ line. The origin corresponds to the position of the quasar which is marked by a `Q', whereas the companion galaxy is labeled with a 'C' and is shown in insets for the two galaxy pairs with the furthest angular separation (J0842$+$1218 and J2100$-$1715). The mean velocity field is relative to the redshift of the quasar (see Table \ref{tab:obsfirprop}), except for the insets, where the zero velocity corresponds to the redshift of the companion galaxy in order to highlight the velocity structure of the companion. In all panels, the ALMA beam is plotted in the bottom left. Contours start at 3$\sigma$ and increase by powers of $\sqrt{2}$, negative contours are dashed. Enlarged versions of each individual galaxy pair are given in Appendix \ref{sec:ApxMomMap}, which also shows and inset for the second companion galaxy of J0842$+$1218.
\label{fig:AllMom}}
\end{figure*}

\section{Notes on Individual Sources}
\label{sec:IndSor}
In this section we provide a detailed description of the individual sources. For each source, the \CII\ spectra is shown in Figure \ref{fig:CIISpec}, whereas the continuum, integrated \CII\ line, velocity and velocity dispersion maps of all of the sources are shown in Figure \ref{fig:AllMom}. Larger images of the individual sources are given in Appendix \ref{sec:ApxMomMap}, while channel maps of all of the individual sources are given in Appendix \ref{sec:ApxChanMap}. Finally, the observed far-infrared properties of the sources are given in Table \ref{tab:obsfirprop}, and the derived properties from these observations are tabulated in Table \ref{tab:derfirprop}. 

The observed and derived properties are obtained as follows. The position on the sky (R.A. and Dec.), as well as the area of the emission, are obtained from fitting a 2D-Gaussian to both the \CII\ and dust continuum emission using the task \texttt{imfit} in CASA. These positions were then used to measure the impact parameters between the quasars host and companion galaxies. The continuum and \CII\ flux densities were estimated from a $1\farcs5$ diameter region centered on the emission. This size was chosen because the lower resolution data showed no significant emission beyond this region. The redshift and full width at half maximum (FWHM) of the \CII\ line are derived from a Gaussian fit to the spectrum derived from this region. 

The \CII\ luminosities are derived from the \CII\ fluxes using standard equations \citep[e.g.][]{Solomon1997,Carilli2013}. The total infrared luminosity, \LTIR, was estimated from the continuum flux measurement by fitting a modified black-body spectrum to the data \citep[see e.g.,][]{Venemans2016}. We vary the dust temperature ($T_d = 30 - 50$~K) and spectral index ($\beta = 1.2 - 2.0$) of the modified black-body spectrum as these quantities are unconstraint by the single data point. The resultant range of total infrared luminosities are given in Table \ref{tab:derfirprop}. SFRs are estimated using two approaches. First, using the relationship between SFR and \CII\ luminosity \citep{Delooze2014,Herrera-Camus2015}, where the quoted range in SFR includes a 0.5 dex uncertainty arising from the scatter around this relationship. Second, using the conversion between SFR and dust continuum measurement / total infrared luminosity \citep[e.g.,][]{Kennicutt2012}. Finally the dynamical masses are estimated using the \CII\ FWHM measurement as a proxy for the circular velocity, and from modeling the \CII\ emission, both methods are described in detail in Section \ref{sec:mass}.

\subsection{J0842$+$1218}
From the initial observations discussed in \citet{Decarli2017}, we know that quasar J0842$+$1218 has a bright \CII-emitting companion galaxy 47~kpc northwest of the position of the quasar. The higher resolution \CII\ observations detect both the quasar host and this companion galaxy at high significance. The \CII\ spectrum for both sources are shown in Figure \ref{fig:CIISpec}A. The integrated \CII\ flux densities of $1.40 \pm 0.14$ and $1.80 \pm 0.24$~Jy~\kms\ and redshifts of $z=6.0760$ and $z=6.0656$ for the quasar host and companion galaxy, respectively, are fully consistent with the previous observations \citep{Decarli2017, Decarli2018}. Even in these deeper observations, we see little evidence for any deviation from a Gaussian profile, both the \CII\ spectrum of the quasar host and companion galaxy can be accurately modeled by a single Gaussian fit. However, the deeper observations do reveal an additional \CII\ emitter southeast of the quasar at an impact parameter of $5\farcs4$ (31~kpc). The integrated \CII\ flux density of this emitter is $0.41 \pm 0.09$~Jy~\kms\ and is offset in velocity by $-480$~\kms\ compared to the quasar host galaxy. This is similar to the velocity offset of the brighter companion galaxy ($-460$~\kms).

The 260~GHz continuum image for the field surrounding quasar J0842$+$1218 yields a dust continuum flux density for  the quasar host galaxy of $0.72 \pm 0.06$~mJy, consistent with the flux measurement from the previous observations. The emission is marginally resolved at this resolution. Using \texttt{imfit} in CASA, we measure the extent of the continuum emission, after deconvolution with the beam, as $(0\farcs28 \times 0\farcs16)$. We note that since this routine fits a 2D Gaussian to the emission, the size might not capture low-level non-Gaussian extended emission. This is discussed in detail in Section \ref{sec:Size}. These deeper observations further yield a $0.23 \pm 0.05$~mJy (4.6$\sigma$) detection of the continuum of the companion galaxy. However, this faint emission remains unresolved in these observations. The second, weaker \CII\ companion remains undetected in the continuum image.

From the continuum-subtracted, integrated \CII\ flux density (second column in Fig. \ref{fig:AllMom}), we can measure the extent of the \CII\ emission. As with the continuum emission, the \CII\ line remains very compact with marginally resolved sizes of $(0\farcs57 \times 0\farcs37)$ and $(0\farcs48 \times 0\farcs30)$ for the quasar host and companion galaxy, respectively. This corresponds to physical sizes of $\approx$3~kpc at the redshift of the galaxies. The fainter, second companion is not resolved in these images. Although the extent of the \CII\ emission between quasar host and companion galaxy is similar, when we generate velocity and velocity dispersion maps of the \CII\ emission (third and fourth columns of Fig. \ref{fig:AllMom}), the companion galaxy shows a velocity gradient along an axis with a position angle of $197 \pm 8 \degr$ east of north (see Sec. \ref{sec:mass}). Such a velocity gradient could be an indication of rotation. However, the marginally resolved observations cannot rule out other scenarios, such as two merging clumps, which would give a similar velocity gradient. No velocity gradient is seen in the quasar host galaxy.

\subsection{PJ167$-$13}
The discovery \CII\ spectra of quasar PJ167$-$13 showed no conclusive evidence for a companion in the field \citep{Decarli2018}. However, deeper and better resolution ($\approx$0.7$\arcsec$) observations of the quasar revealed a clear excess of \CII\ and dust emission $0\farcs92$ southeast of the quasar \citep{Willott2017}. This emission was attributed to a close companion at a separation of only 5~kpc from the quasar host galaxy. This companion source remains the only source of our sample that has been detected in the optical/near-infrared \citep{Mazzucchelli2019}. Our higher resolution data confirms the existence of a distinct \CII\ and dust continuum emitter at this position. By extracting \CII\ spectra for both the quasar host and the companion galaxy, individually, we find that both spectra are accurately described by a Gaussian profile (Fig. \ref{fig:CIISpec}B). Unlike the results in \citet{Willott2017}, the addition of the flux from both the quasar host and the companion galaxy results in an accurate fit of the total flux of the complex (yellow line in Fig. \ref{fig:CIISpec}B). No additional source of emission is needed. The discrepancy between the two results is due to the slightly higher redshift determination of the companion galaxy in this work ($-140$~\kms, instead of $-270$~\kms\ relative to the peak \CII\ emission from the quasar host). This more accurate redshift determination was enabled by the improved spatial separation between the sources.

The dust continuum and \CII\ integrated flux map for PJ167$-$13 show that the companion galaxy can be clearly distinguished as a separate density peak (Fig. \ref{fig:AllMom}). There is also evidence for significant emission arising from gas that lies between the quasar host and companion galaxy. This gas has a mean velocity between the systemic velocity of the quasar host and companion galaxies, resulting in a smooth velocity gradient for the quasar host -- companion galaxy pair. This indicates that the excess gas is smoothly distributed between the two galaxies, likely the result of tidal interactions between the two interacting galaxies. The high velocity dispersion throughout the system indicates the emitting gas is turbulent, which is expected for gas inside a merging system.

\subsection{J1306$+$0356}
Previous $\sim$1$\arcsec$ \CII\ observations of the host galaxy of quasar J1306$+$0356 revealed that the \CII\ emission is extended over a $1\farcs43 \times 0\farcs74$ region \citep{Decarli2018}. No significant velocity gradient was evident in these observations. The current higher resolution observations reveal that the \CII\ emission arises from two spatially and spectrally distinct sources with a physical separation of 5.4~kpc and velocity separation of 60~\kms. Both \CII\ emission features can be accurately described by a Gaussian fit (Fig. \ref{fig:CIISpec}C). We interpret this emission as arising from two distinct galaxies with integrated fluxes of  $1.80 \pm 0.15$ and $0.86 \pm 0.11$~Jy~\kms\ for the quasar host and companion galaxy, respectively. Both sources are also detected in the 263~GHz continuum observations ($0.92 \pm 0.09$ and $0.30 \pm 0.09$~mJy). As with PJ167$-$13, excess gas is observed between the companion and quasar host, suggesting that these galaxies are actively interacting.

The \CII\ emission from both galaxies appears compact, with deconvolved sizes for the quasar host and companion galaxy of $0\farcs56 \times 0\farcs42$ and $0\farcs64 \times 0\farcs49$, respectively. In addition, the velocity field of the two galaxies suggests little coherent motion in either source, although the companion galaxy shows a possible velocity gradient along the north-east to south-west direction. The excess gas between the galaxies has a velocity that is most similar to the companion galaxy, suggestive of a gaseous tidal feature associated with the companion galaxy due to an interaction with the quasar host. Velocity dispersions in the quasar host and companion galaxy are similar.

\subsection{PJ231$-$20}
The \CII\ discovery spectra taken of the field surrounding PJ231$-$20 revealed a strong \CII\ emitter at 9.1~kpc distance from the quasar host. The new observations yield continuum and integrated line fluxes that are consistent with the results previously obtained \citep{Decarli2018}. In addition to these two sources, the deeper observations reveal an additional, weaker, \CII\ emitter 14~kpc south-southeast of the quasar, 6~kpc from the companion galaxy. It has an integrated flux of $0.26 \pm 0.07$~Jy~\kms, and is redshifted by 400~\kms\ compared to the redshift of the quasar host. A slight continuum excess is also seen at this position, resulting in a far-infrared continuum flux measurement for this source of $0.19 \pm 0.04$~mJy (Fig. \ref{fig:CIISpec}D).

The continuum emission, integrated \CII\ flux, velocity and velocity dispersion maps of this galaxy pair are shown in the fourth row of Figure \ref{fig:AllMom}. Unlike the lower resolution data, the emission from the quasar host and companion galaxies are clearly separated into two distinct sources without significant overlap. A Gaussian 2D fit to the data reveals that the emission from both sources is compact. However, faint, much more extended, emission surrounds both sources. This suggests some gas has already been stripped from the galaxy due to gravitational interactions. It is therefore very likely that these galaxies are in the beginning stages of actively interacting with each other. In this scenario, the second \CII\ emitter east of the companion galaxy could be dense gas stripped from the companion by tidal forces.

In addition to the similar velocity fields, we also see little difference between the extent of the continuum and/or \CII\ line emission between quasar host and companion galaxy. Both objects have roughly equal sizes (see Section \ref{sec:Size}), providing no obvious clues why one of the galaxies hosts an unobscured quasar and the other does not.

\subsection{J2100$-$1715}
The largest separation quasar host -- companion galaxy pair of the sample, the quasar host of quasar J2100$-$1715 and its companion galaxy have a separation of $10.8\arcsec$ (61~kpc). As Figure \ref{fig:CIISpec}E shows, both the quasar host and the companion galaxy are bright, with fluxes of $2.26 \pm 0.22$ and $4.15 \pm 0.52$~Jy~\kms, respectively. These fluxes are consistent with the previous observations after taking into account the extent of the emission \citep{Decarli2018}. The \CII\ spectra of the companion galaxy is the only spectrum in our sample that shows a deviation from a Gaussian profile, the emission is flattened compared to the best-fit Gaussian shown in Figure \ref{fig:CIISpec}E. We also confirm the detection of the continuum emission for both sources, which have fluxes of $0.88 \pm 0.09$ and $2.25 \pm 0.21$ mJy.

In addition to the non-Gaussian \CII\ profile, the companion galaxy shows a strong velocity gradient along an axis with a position angle of 13$\degr$. This is in contrast with the quasar host galaxy, which shows little ordered motion. The non-Gaussianity of the companion galaxy's \CII\ profile is likely driven by this strong velocity gradient. Such a velocity gradient is consistent with the \CII\ emission arising from gas rotating in a disk. This is further corroborated by the elongated extent of the \CII\ emission ($0\farcs66 \times 0\farcs31$) and the kinematic modeling (Sec. \ref{sec:mass}). However, the resolution is insufficient to rule out other possible scenarios. The \CII\ emission and dust continuum extent of the quasar host are similar to the companion galaxy, both are listed in Table \ref{tab:obsfirprop}.

\begin{figure*}[!t]
\includegraphics[width=\textwidth]{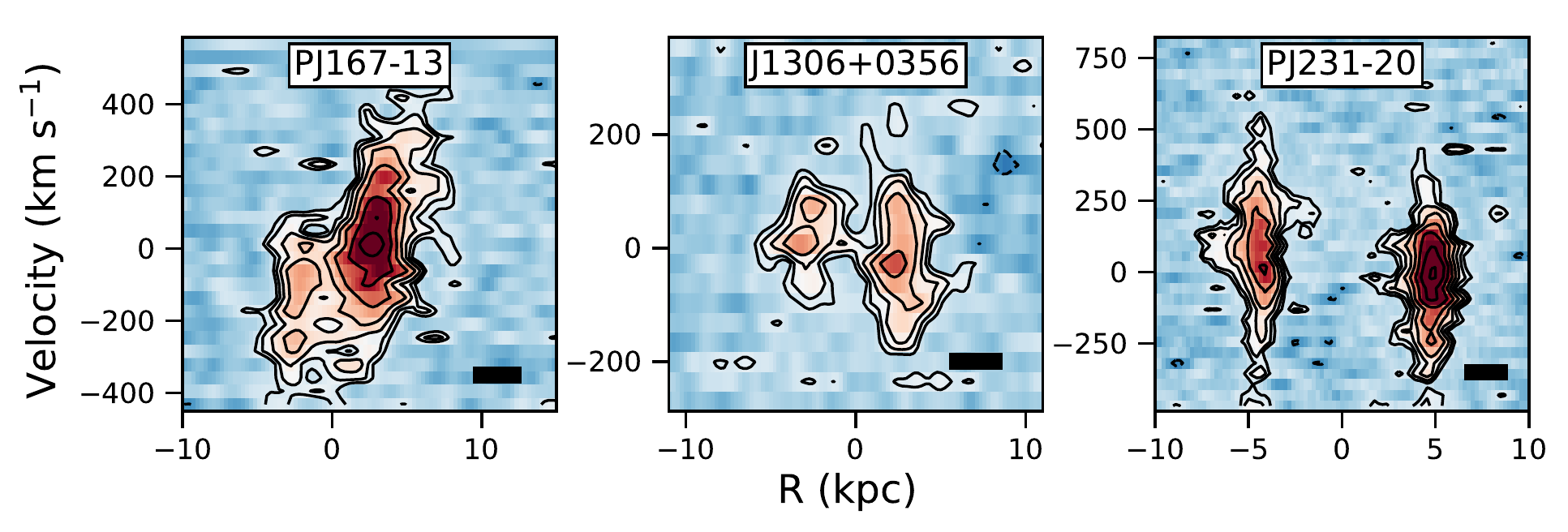}
\caption{Position-velocity ($p$-$v$) diagrams of the three closest quasar host -- companion galaxy pairs. These $p$-$v$ diagrams were generated from the line that intersects the centers of both galaxies. In all panels the companion galaxies are to the left of the quasar host. Outer contours are at 2$\sigma$ and increase by powers of $\sqrt{2}$ (negative contours dashed). The size of the ALMA synthesized beam is displayed by a black horizontal bar in the bottom right corner. The left two galaxy pairs show evidence of \CII\ emitting gas connecting the two galaxies. This gas is also seen in dust continuum emission.
\label{fig:PV}}
\end{figure*}

\section{Analysis}

\subsection{Emission Surrounding Interacting Systems}
Galaxies that are interacting are expected to show gas distributions that are more extended and perturbed due to tidal forces during the encounter. This is seen at low redshift in abundance, and has also been observed at high redshift in dust continuum \citep{Diaz-Santos2018}. In the two closest-separation quasar host -- companion galaxy pairs discussed in this manuscript, a clear excess of gas, which is detected both in continuum and in \CII\ emission, can be seen connecting both galaxies (Fig. \ref{fig:AllMom} second and third row). The third-closest galaxy pair (PJ231$-$20) shows a perturbed gas distribution with significant faint, extended \CII\ emission both between the galaxies and in the immediate environs. To highlight the gas between the galaxies, we generate a position-velocity ($p$-$v$) diagram that is oriented such that it intersects both galaxies for these three galaxy pairs in Figure \ref{fig:PV}. 

This figure shows that the gas connecting the host of QSO PJ167$-$13 and its companion galaxy in projection, also shows a smooth velocity gradient from one galaxy to the other. This is reminiscent of the smooth velocity gradient of the nearby galaxy merger between M81 and M82 \citep[e.g.,][]{DeBlok2018}. The emission between the galaxy pair toward J1306$+$0356 is more tenuous, and seems to have a kinematic profile that is more consistent with the companion galaxy than the quasar host. For PJ231$-$20, no direct gas connection is detected. However, both the quasar host and companion galaxies show perturbed gas distributions that are both more extended (see Section \ref{sec:Size}) and irregular compared to the gas distributions of the galaxies forming the two wide separation pairs. This suggests gravity has already perturbed the gas distribution in this galaxy pair.

\subsection{Size Estimates and Extent of Emission}
\label{sec:Size}
\begin{figure*}[!t]
\includegraphics[width=\textwidth]{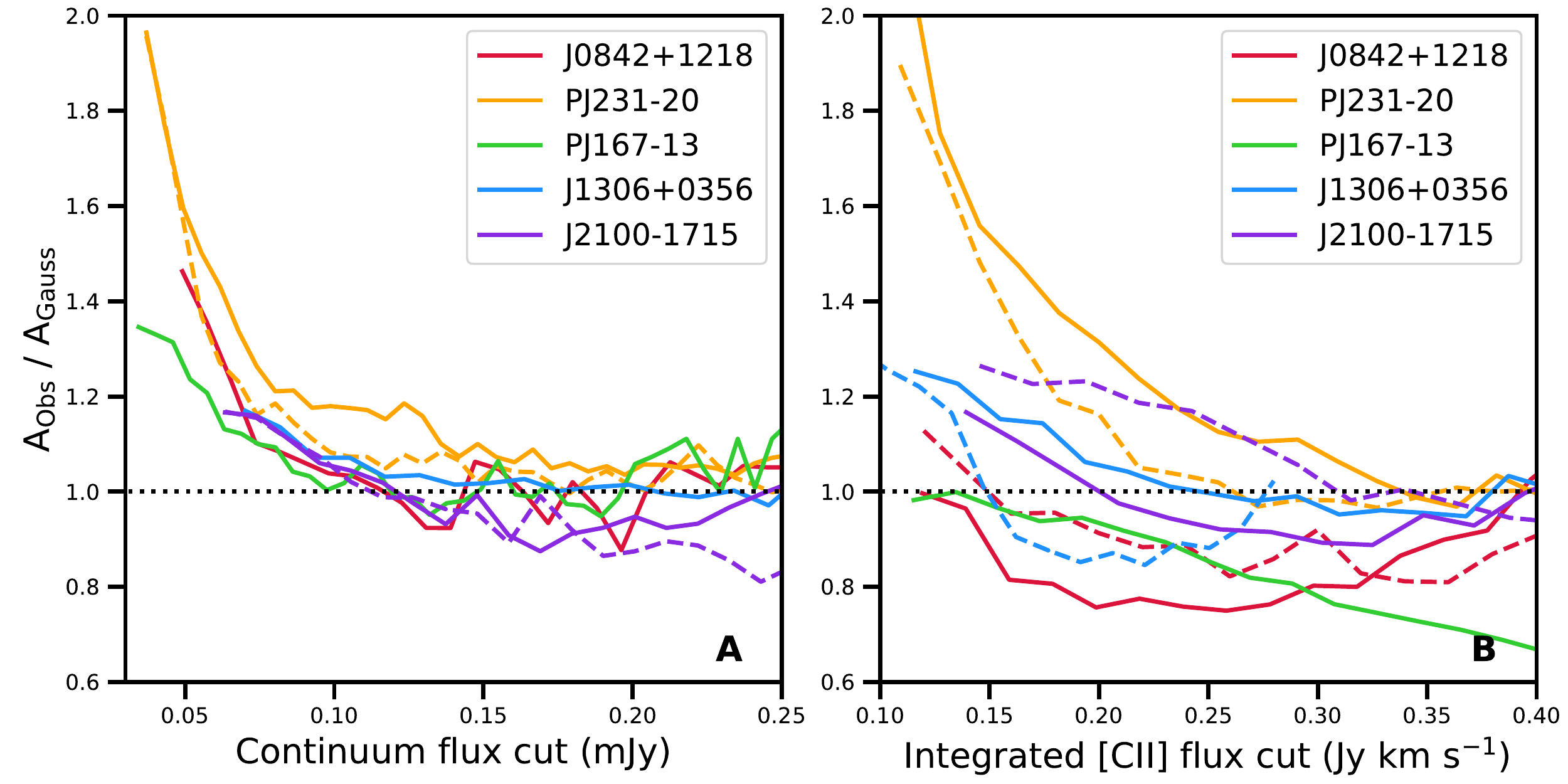}
\caption{Size ratio of the area of the observed emission (A$_{\rm obs}$), compared to the size of the fitted 2D Gaussian (A$_{\rm Gauss}$) as a function of the applied flux density cut for the continuum flux (left panel) and the integrated \CII\ flux (right panel). Deviations from unity imply that the 2D gaussian fit is not accurately describing the observed flux. Lines for both the quasar host (solid lines) and companion galaxies (dotted lines) start at the $3\sigma$ flux cut for that object, and increase in steps of 3$\sigma$. The largest deviation from unity occurs for small flux cuts surrounding the quasar host and companion galaxy of PJ231$-$20. This is an indication that these galaxies have faint extended emission that is not recovered by the 2D Gaussian fitting routine. The smaller-than-unity ratio in the integrated \CII\ flux for the quasar host of PJ167$-$13 suggests this galaxy's \CII\ profile is flatter than a 2D Gaussian, which is likely the result of the gravitational interaction with its nearby companion galaxy.
\label{fig:Sizes}}
\end{figure*}

The size estimates in Table \ref{tab:obsfirprop} are based on the assumption that the emission can be accurately described by a 2D Gaussian. For resolved observations of objects with non-Gaussian surface brightness profiles, this might not be a valid assumption. If, for instance, there exists low-level extended emission around a compact bright source, the fitting routine might not accurately describe the extent of the low-level emission as the fit is dominated by the compact source. To ascertain if such low-level flux exists in these resolved observations, we fit a 2D Gaussian to the total emission for each of our sources. We then compare the area of the emission for this 2D Gaussian ($A_{\rm Gauss}$) to the area of the observed emission ($A_{\rm obs}$) for a range of different flux values. Here, both areas areas remain convolved with the ALMA beam. The results are plotted in Figure \ref{fig:Sizes}. If the shape of the emission was a pure 2D Gaussian, each line would have a constant value of 1 independent of the flux cut.

We can see that for the strongest emission of the continuum flux density (flux cuts $> 0.1$~mJy) the ratio of the Gaussian size to the observed size is $\sim 1$. This implies that this emission must come from a compact source that can be accurately described by a 2D Gaussian. However at lower continuum flux cuts, the emission is more extended than what is predicted from a Gaussian shape. This deviation from a Gaussian shape could be caused by calibration issues, or it could be actual low-level emission that is not modeled by the Gaussian fitting routine. As the largest deviations occur for PJ231$-$20, two close, interacting galaxies, we posit that, at least for this system, the low-level emission arises from more extended gas, and the size/area estimates from the Gaussian routine could be off by as much as a factor of two.

The same analysis can be done on the integrated \CII\ line emission map. The results are shown in Figure \ref{fig:Sizes}B. Again the two largest deviators in the plot are the quasar host and companion galaxy of PJ231$-$20, indicative of low-level \CII\ emission from gas on more extended scales around these galaxies. The remaining systems show observed sizes that are roughly consistent with the Gaussian estimates, although with substantial scatter. This suggests that, unlike the continuum observations, the emission is less centrally concentrated and the Gaussian fit fully captures the extended \CII\ emission. Indeed, for some emitters --such as the quasar host galaxy of PJ167$-$13-- the Gaussian sizes are actually larger than the observed sizes, suggesting that these emitters have a surface brightness profile which is flatter than a 2D Gaussian.

\subsection{[C${\scriptscriptstyle{II}}$] Deficit}
\label{sec:CIIdef}
A well-known property of both low and high redshift galaxies, is the decrease in \CII\ to total infrared (TIR) luminosity as the TIR luminosity of the galaxy increases. The cause for this deficit in \CII\ luminosity for more TIR-luminous systems is still debated. One possibility is that the TIR luminosity is enhanced in TIR-luminous galaxies, because of an increased UV radiation field, either due to the presence of an active galactic nucleus or due to increased star formation \citep{Malhotra1997}. Other possibilities suggest that \CII\ emission is suppressed in TIR-luminous systems, either because of dust absorption of the \CII\ line \citep{Riechers2013, Riechers2014}, or because of intrinsic properties of the gas responsible for the \CII\ emission \citep{Munoz2016, Narayanan2017}. Observationally, galaxies hosting an active galactic nucleus (AGN) or are actively interacting have the highest \CII\ deficits \citep[e.g.,][]{Sargsyan2012, Carilli2013, Farrah2013}. Apportioning our sample by type and impact parameter, allows us to see if these observational trends hold for our current sample as well. In addition, we can observe for a spatial variation in the \CII\ deficit for the individual sources, as most sources are resolved.

\begin{figure*}[!t]
\includegraphics[width=\textwidth]{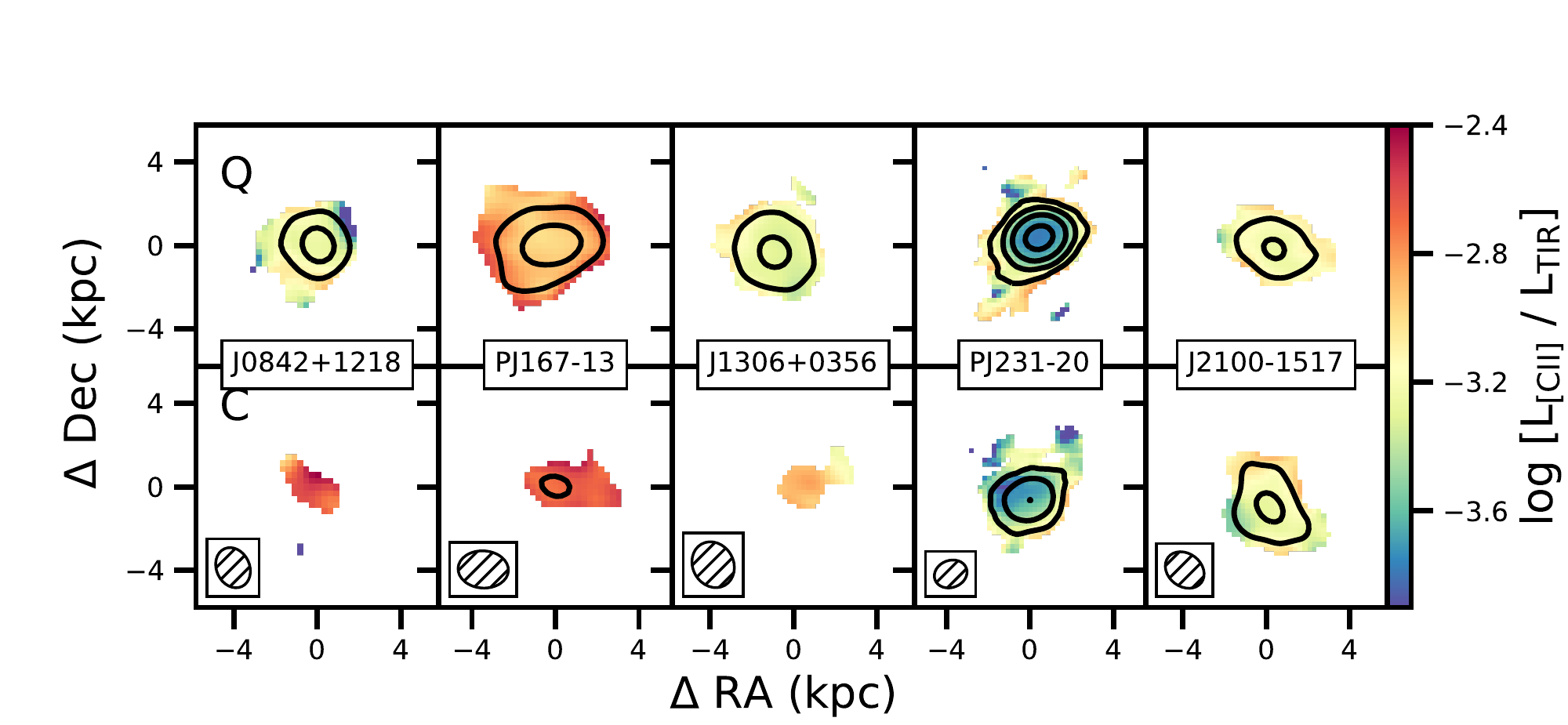}
\caption{Spatially resolved \CII\ deficit for the quasar host (\emph{top row}) and the companion (\emph{bottom row}) galaxies. Contours show surfaces of constant TIR luminosity surface density. Within these contours, the  \LCII/\LTIR\ ratio remains roughly constant. All galaxies show elevated \CII\ deficits in the center of the galaxy, because the continuum emission is generally more compact compared to the \CII\ emission. No significant differences in  \LCII/\LTIR\ ratio are seen between the quasar host and companion galaxies, and the range of \CII\ deficits probed is typical for TIR-luminous galaxies \citep{Diaz-Santos2017}. The synthesized beams for the observations are shown in the bottom row in the bottom left inset.
\label{fig:CIIDef}}
\end{figure*}

\begin{figure}[!t]
\includegraphics[width=0.48\textwidth]{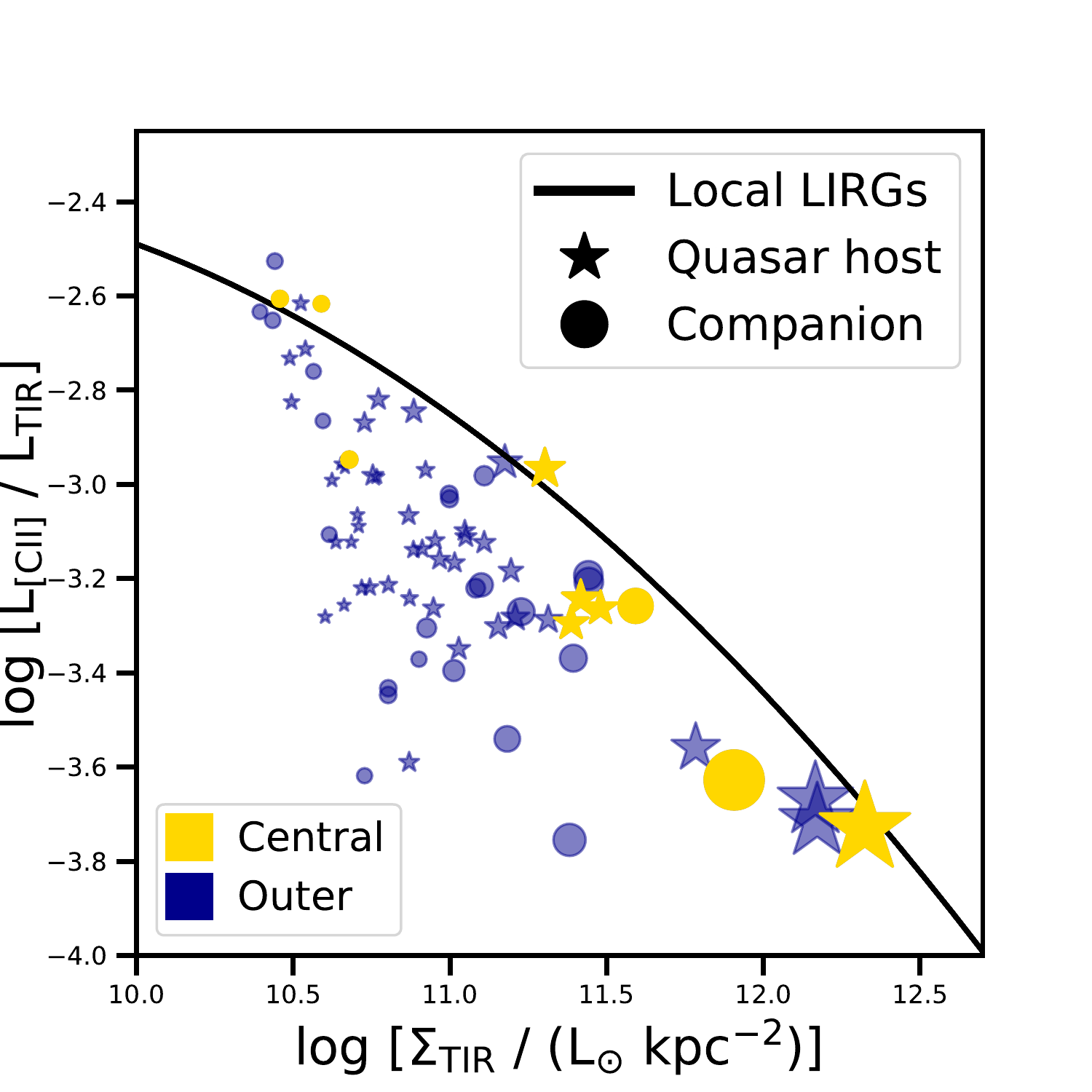}
\caption{Spatially resolved \CII\ deficit as a function of the TIR luminosity surface density. The size of the markers is scaled to the signal-to-noise ratio of the continuum measurement. The data was sampled at 0$\farcs$3 resolution, giving $1-2$ measurements per independent beam for each observation. Only measurements where the continuum was detected at greater than 3$\sigma$ are shown. For each source the value of the \CII\ deficit at the peak of the continuum flux is marked by a yellow symbol. These values are in excellent agreement with the relationship for local luminous infrared galaxies \citep[LIRGs -- black line;][]{Diaz-Santos2017}}
\label{fig:FIRCII}
\end{figure}

Figure \ref{fig:CIIDef} shows the spatial extent of the \CII\ deficit for all sources. This figure highlights that brighter sources have lower \CII-to-TIR luminosity ratios, \LCII/\LTIR. This is further exemplified by plotting the \CII\ deficit as a function of TIR luminosity surface density, $\Sigma_{\rm TIR}$ for the sources (Figure \ref{fig:FIRCII}). The yellow symbols mark the values for the central pixel of each source. If quasars (i.e., AGNs) are the dominant cause of the \CII\ deficit, we would expect to see a difference between the quasar host and companion galaxies. However, the central pixels for both quasars host and companion galaxies show remarkable agreement with a compilation of \CII\ deficit measurements from local infrared-bright galaxies \citep[LIRGs;][]{Diaz-Santos2017}. Unless all of the companion galaxies host an obscured AGN, which we consider unlikely (see Sec. \ref{sec:comp}), the presence of an AGN in the quasar host galaxies does not add to the \CII\ deficit. Similarly, we find no evidence for a decreased \LCII/\LTIR\ in actively interacting galaxy pairs. Although the scatter in  \LCII/\LTIR\ does seem to increase in these systems, the average is comparable to the average of the large impact parameter galaxy pairs.

Figure \ref{fig:CIIDef} further shows that the \CII\ deficit is relatively constant within the inner part of the galaxies. This is similar to the spatially-resolved mergers observed at high redshift \citep[e.g.,][]{Neri2014, Litke2019, Rybak2019}. The constant ratio can be explained if the bulk of \CII\ and TIR emission arises from a compact source, which is consistent with the result in section \ref{sec:Size}. For nearly all galaxies the \CII\ deficit gets smaller toward the edges of the galaxy. However as Figure \ref{fig:FIRCII} shows, the outer edges still follow the global trend of larger \CII\ deficits for increasing TIR luminosity surface densities. The offset between the central and outer parts of the galaxies in this trend could be due to varying physical conditions within these regions, which would corroborate the assertion that \LCII/\LTIR\ is set by local ISM conditions \citep{Munoz2016, Smith2017, Gullberg2018, Herrera-Camus2018}. We note two important caveats to these results: i) These results hold at the resolution of the observations. Any variations in \LCII/\LTIR\ at sub-kpc scale is not resolved by these observations \citep[see e.g.,][]{Venemans2019}. ii) To calculate the TIR luminosity, we assume that the physical conditions of the gas do not vary across the galaxy and are equal to the fiducial values (see Section \ref{sec:IndSor}). This is likely an oversimplification on small scales. 

\subsection{Constraints on Dynamical Mass}
\label{sec:mass}
Several estimators are used in the literature to obtain the dynamical mass of galaxies with unresolved or marginally resolved far-infrared lines. Under the assumption that the gas is rotationally supported, the enclosed mass (in $M_\odot$) within an emission region is $M_{\rm dyn} = 1.16 \times 10^5~v_{\rm circ}^2~D$ \citep[e.g.,][]{Walter2003, Wang2013}. Here, $D$ is the size (diameter) of the emission in kpc and $v_{\rm circ}$ is the circular velocity in \kms. The circular velocity estimate relies on an additional assumption on the kinematics and distribution of the gas. If the gas is virialized but shows non-ordered, dispersion-dominated motion, then $v_{\rm circ}$ is best characterized by the velocity dispersion ($\sigma_v$) of the gas \citep[$\sqrt{3/2}\sigma_v$; e.g.,][]{Decarli2018}. However, if the gas shows order motion, then the full width at half maximum (FWHM) of the \CII\ line can be used as a proxy for the velocity of the system: $v_{\rm circ} = 0.75 \times {\rm FWHM_{[C{\scriptscriptstyle II}]}} / \sin{i}$, where the inclination is often taken to be $55\degr$ \citep[e.g.,][]{Wang2013,Decarli2018}. The range of dynamical mass estimates inferred from applying these two methods are given in Table \ref{tab:derfirprop}.

The higher resolution of the observations presented here allows us to better constrain the dynamical properties of the gas. To accomplish this, we use a custom, python-based code which fits the three-dimensional data cube to a model data cube generated from a user-defined model. The code generates a model data cube with the same spectral and spatial resolution as the data from the user-defined model. It then convolves this data cube with the ALMA synthesized beam, which is compared with the observed data cube using a simple $\chi^2$ statistic. To account for the large spatial correlation between adjacent pixels, the comparison is done using a bootstrap method. Finally, the full parameter space is sampled through a Markov Chain Monte Carlo method by employing the \texttt{emcee} package \citep{Foreman-Mackey2013}. 

In this paper, we model the emission with two different models, one in which the \CII\ emission is due to a purely dispersion-dominated gas, and one in which the \CII\ is emitted from gas in a thin disk with constant circular velocity and constant velocity dispersion. Both models assume that the intensity of the \CII\ line can be modeled by an exponential function, and are describe in more detail in Appendix \ref{sec:ApxModel}. Results from the fitting procedure are shown in Table \ref{tab:kinfit}. 

From this analysis, we conclude that three of the companion galaxies (J0842$+$1218, PJ231$-$20 and J2100$-$1715) and one quasar host galaxy (PJ167$-$13) have \CII\ emission that is consistent with arising from a rotating disk. However, in nearly all cases the emission is highly turbulent, with dispersion velocities roughly equal to the circular velocities. This could indicate that the assumed constant velocity profile is incorrect, because the \CII\ emission is still probing the rising part of the rotation curve \citep{DeBlok2014}. To assess how this affects the circular velocity estimate, we also run a model where the velocity is described by a linearly increasing function. We find that for this model the resulting circular velocity estimates at the maximal extent of the emission are similar to the previous estimates. Therefore the choice of velocity profile does not significantly affect the results of the fitting. For the remaining objects, either the circular velocity must be substantially smaller than the dispersion (e.g., the quasar host galaxy towards PJ231$-$20), and/or the inclination of the galaxies is small.

The large velocity dispersion estimates provide an explanation for the remarkable near-Gaussian shape of the total integrated \CII\ flux spectra \citep[see Fig. \ref{fig:CIISpec} and][]{Decarli2018}. Even for clearly distinct \CII\ emitters (i.e., PJ167$-$13 and J1306$+$0356), the combined \CII\ spectrum of the two sources remains nearly Gaussian, since the large velocity dispersion, compared to the small offset in central frequency, hinders spectral separation. The high velocity dispersions could be caused by turbulent ISM conditions \citep[e.g.,][]{DeBreuck2014}, but could also be the result of sampling the \CII\ emission from the rising part of the rotation curve \citep{DeBlok2014}. In either case, these velocity dispersions are comparable to the velocity dispersions found for $z \sim 2$ compact star-forming galaxies \citep{Barro2014}. An important caveat to these results is that velocity structures on sub-kpc scales, below the resolution of the observations, are smoothed out, causing an increase in the velocity dispersion. In addition, the resolution is insufficient to rule out other possible scenarios, such as merging clumps.

To estimate the dynamical mass for the four systems that have a kinematic signature consistent with a rotating disk, we take the circular velocity estimate from the model, and add in quadrature $\sqrt{3/2}$ times the velocity dispersion estimate. This accounts for part of the velocity dispersion arising from the simplified assumption that the gas is constrained to a thin disk. For the remaining systems, we approximate the circular velocity by $\sqrt{3/2}$ times the velocity dispersion as given by dispersion-dominated model. This will give a lower limit to the dynamical mass. The dynamical masses for all systems range between ($0.3$ $-$ $>$$5.4) \times 10^{10} M_\odot$ (Table \ref{tab:derfirprop}). 

The dynamical mass estimates obtained through kinematic modeling are in rough agreement with the estimates obtained using the \CII\ line profile. For the two systems with well constrained dynamical mass estimates (PJ167$-$13Q and J2100$-$1715C), the kinematics modeling estimates are on the lower end of the dynamical range estimate obtained from the \CII\ line profile. This is due to the the large velocity dispersion in these systems, which widens the line profile. Thus, for systems with large velocity dispersions, using the FWHM of the \CII\ line as a proxy for the circular velocity could overestimate their dynamical mass estimate. Finally we note that standard practice has been to assume an inclination angle of 55$\degr$ for quasar host galaxies \citep{Wang2013,Willott2015,Decarli2018}. We can rule out this inclination for four out of the six quasar hosts, and derive a mean inclination of $<$$39\degr$.

\subsection{QSO Host and Companion Galaxy Comparison}
\label{sec:comp}
One of the primary aims of this study is to compare the far-infrared properties of the quasar host galaxies to the far-infrared properties of the brightest companion host galaxies in order to examine the effect, if any, the central accreting supermassive black hole has on these properties. As shown in the previous sections, we find no evidence that the quasar affects the strength or extent of either the \CII\ or continuum emission. In addition, the \LCII/\LTIR\ ratio of both galaxy populations is similar (Fig. \ref{fig:FIRCII}). These observations therefore corroborate the assertion that the central accreting supermassive black holes in quasars do not significantly alter the observed \CII\ and dust continuum emission, instead this emission originates predominantly from heating of the ISM by stars \citep[e.g.,][]{Venemans2017}. This is in agreement with the [O\,\textsc{iii}]~88~$\mu$m observations of one of the galaxy pairs (J2100$-$1715), which revealed little difference in the far-infrared emission properties of the ionized gas \citep{Walter2018}.

A possible caveat to this result is that the companion galaxies could host an obscured AGN. If some fraction of companion galaxies host an obscured AGN which significantly alter their far-infrared properties, we would expect to see a bimodal distribution within the companion galaxies' far-infrared properties. The lack of such a bimodality implies that either none or all of the companion galaxies host an obscured AGN. If we assume the latter and assume an AGN obscuration factor of $\sim$50\% \citep{Merloni2014}, then the likelihood of not detecting any quasar-quasar pairs (i.e., where both AGNs are optically unobscured) is 3\%. Formally this is an upper limit, as no close quasar-quasar pairs are known at $z > 6$, whereas several non-quasar host, far-infrared luminous galaxy pairs are known at high redshift \citep{Oteo2016,Riechers2017,Marrone2018}. We therefore disfavor this scenario, and conclude that quasars do not significantly alter the \CII\ and far-infrared continuum emission measurements of high redshift galaxies.

In these resolved \CII\ observations, there is a possible hint that the kinematics or orientation of the companion hosts are slightly different, as three out of five companion galaxies show signs of rotation, compared to only one out of five quasar host galaxies. One possible explanation for such a difference is that the gas kinematics in the quasar hosts are more disturbed due to previous mergers unrelated to the current merger. In this scenario, the quasar hosts are still in a post-merger state characterized by perturbed gas kinematics, whereas the companion galaxies have not experienced a recent merger, and show more ordered rotation. Another possibility is that we are preferentially observing the quasar host galaxies face-on, thereby minimizing any velocity gradient caused by rotation. Such an orientation is at least consistent with the observation of a bright luminous quasar in a dusty galaxy. However, the sample remains too small to confirm potential differences of this magnitude in the kinematics of the gas.

\section{Summary and Conclusions}
One of the most surprising results from the \CII\ emission study of $z > 6$ quasars has been the high rate of strong \CII\ emitting companion sources surrounding these quasars. Of the 27 quasars targeted in \CII\, four show clear evidence of a distinct companion source \citep{Decarli2017, Decarli2018}. Higher resolution observations reveal that two additional quasars, previously identified as being `extended', have a nearby companion galaxy \citep[and this work]{Willott2017}. These \CII-bright companion galaxies have far-infrared properties similar to the quasar hosts, but lack an extremely luminous AGN at its center. This paper discusses resolved ($\approx$$0\farcs35$; $\approx$2~kpc) \CII\ observations of five out of the six systems, while the quasar host companion galaxy pair toward PJ308$-$21 is discussed in \citet{Decarli2019}.  The results are:

\begin{itemize}
\item{All ten sources (i.e., five quasar-host/companion pairs)  are detected in both \CII\ emission and continuum emission. These measurements are within the uncertainties consistent with the lower resolution data \citep{Decarli2017, Decarli2018} in which not all sources were detected in continuum. The companion galaxies are bright in the far-infrared with roughly similar luminosities as the quasar host. However, both the quasar hosts and companion galaxies show a large range in observed fluxes and resulting luminosities (Table \ref{tab:obsfirprop}).}

\item{The two closest-separation quasars (PJ167$-$13 and J1306$+$0356) show \CII\ and dust continuum emission that connects both galaxies. Such a bridge of gas has been seen recently in the dust continuum of a high redshift dusty galaxy \citep{Diaz-Santos2018}. Using the velocity information of the \CII\ emission, we determine that the gas in  PJ167$-$13 is directly linking the two galaxies, whereas the gas bridge in J1306$+$0356 is more closely linked to the companion host galaxy.}

\item{No discernible difference is found in the extent of the emission (both \CII\ and far-infrared continuum) between the quasar host and companion galaxies. In addition, size does not seem to correlate with strength of emission. Even in these higher resolution observations, the \CII\ and far-infrared continuum emission remains very compact with typical sizes $\lesssim$3~kpc. The notable exception is the field surrounding PJ231$-$20 which shows significant extended \CII\ emission, including a third \CII\ emitter 6~kpc from the companion galaxy. We find that a standard 2D Gaussian fit of this source underestimates the true size of the emission by a factor of two. We interpret this faint emission as tidal debris from the gravitational interaction between the two interacting galaxies.}

\item{Both companion host galaxies and quasar host galaxies have \CII\ deficits consistent with the results from studies of local ULIRGs \citep{Diaz-Santos2017}. Comparing the spatial distribution of the \CII\ deficit for each galaxy shows that the \CII\ deficit rises at the edges of the emission. This is due to the smaller size of the dust continuum emission compared to the \CII\ emission. However, the majority of galaxies have a near-constant \CII\ deficit across most of the observed emission region.}

\item{Kinematic modeling shows that four galaxies (three companion galaxies and one quasar host galaxy) show \CII\ emission that is consistent with arising from a disk. The remaining galaxies have kinematic properties that suggest either a face-on geometry or a dispersion-dominated velocity profile. The higher fraction of disk-like emission in companion galaxies could be an indicator that we are preferentially seeing the quasar host galaxies face-on. For all sources we estimate a dynamical mass from the kinematic modeling, which ranges between $7 \times 10^9$ and $>$$9 \times 10^{10}~M_\odot$.} 
\end{itemize}

All these results imply that even though the \CII\ emitting gas is relatively compact ($<$5~kpc) and surrounds the luminous quasar, the quasar does not contribute significantly to either the strength of the \CII\ line, or is kinematically altering the gas. Although companion galaxies appear to show more ordered gas motion (three out of five galaxies), the current sample remains too small to make a definitive statement. The similarity in extent and dynamical characteristics corroborates previous assertions that the quasar does not contribute significantly to the \CII\ and dust continuum emission \citep{Venemans2017}. It therefore remains unclear from these observations what differentiates the quasar host galaxies from the companion galaxies in order for them to host a luminous quasar. Facilitated by the compact \CII\ emission, further higher resolution \CII\ observations of these systems, could provide an answer through a detailed look at the sub-kpc motion of the ISM in the centers of these systems. 

\acknowledgments
We would like to thank the referee for constructive comments that help clarify the paper. This paper makes use of the following ALMA data: ADS/JAO.ALMA \#2015.1.01115.S, \#2016.1.00544.S, and \#2017.1.01301.S. ALMA is a partnership of ESO (representing its member states), NSF (USA) and NINS (Japan), together with NRC (Canada), NSC and ASIAA (Taiwan), and KASI (Republic of Korea), in cooperation with the Republic of Chile. The Joint ALMA Observatory is operated by ESO, AUI/NRAO and NAOJ. M.N., F.W., B.V. and Ml.N. acknowledge support from ERC Advanced grant 740246 (Cosmic{\verb|_|}Gas). D.R. acknowledges support from the National Science Foundation under grant number AST-1614213.

\bibliography{Bib.bib}

\movetabledown=1.5in
\begin{rotatetable*}
\begin{deluxetable*}{lccccccccc}
\tablecaption{Parameters of the kinematic modeling
\label{tab:kinfit}}
\tablehead{
\colhead{Name} & 
\colhead{R.A. ($x_c$)} &
\colhead{Dec. ($y_c$)} &
\colhead{$z_{\rm kin}$} &
\colhead{$v_{\rm circ}$\tablenotemark{a}} &
\colhead{$\sigma_v$\tablenotemark{b}} &
\colhead{$i$\tablenotemark{c}} &
\colhead{$\alpha$\tablenotemark{d}} &
\colhead{$I_0$\tablenotemark{e}} &
\colhead{$R_{\rm d}$\tablenotemark{f}} \\
\colhead{} & 
\colhead{(J2000)} &
\colhead{(J2000)} &
\colhead{} &
\colhead{(\kms)} &
\colhead{(\kms)} &
\colhead{($\degr$)} &
\colhead{($\degr$)} &
\colhead{(mJy~beam$^{-1}$)} &
\colhead{(kpc)}
} 
\startdata
\cutinhead{Thin disk model}
J0842$+$1218Q & 08:42:29.4376(8) & $+$12:18:50.437(14) & $6.07639_{-0.00022}^{+0.00020}$ & ---\tablenotemark{g} & $157_{-8}^{+9}$ & $<$$46$ & $174_{-24}^{+25}$ & $3.9_{-0.5}^{+0.6}$ & $0.83_{-0.09}^{+0.12}$\\ 
J0842$+$1218C & 08:42:28.9743(7) & $+$12:18:54.966(11) & $6.06655_{-0.00017}^{+0.00018}$ & $>$$220$ & $129_{-9}^{+10}$ & $<$$28$ & $197_{-7}^{+8}$ & $5.9_{-0.7}^{+0.9}$ & $0.73_{-0.06}^{+0.06}$\\  
\hline
PJ167$-$13Q & 11:10:33.9829(6) & $-$13:29:45.863(6) & $6.51625_{-0.00010}^{+0.00010}$ & $114_{-10}^{+9}$ & $170_{-4}^{+5}$ & $57_{-2}^{+2}$ & $303_{-2}^{+2}$ & $5.1_{-0.3}^{+0.3}$ & $1.61_{-0.07}^{+0.07}$\\  
PJ167$-$13C & 11:10:34.031(2) & $-$13:29:46.278(18) & $6.5125_{-0.0003}^{+0.0003}$ & ---\tablenotemark{g} & $181_{-12}^{+14}$ & $<$$55$ & $340_{-40}^{+50}$ & $2.7_{-0.3}^{+0.4}$ & $1.17_{-0.15}^{+0.19}$\\
\hline
J1306$+$0356Q & 13:06:08.2648(11) & $+$03:56:26.233(13) & $6.03386_{-0.00014}^{+0.00013}$ & ---\tablenotemark{g} & $106_{-6}^{+7}$ & $<$$55$ & ---\tablenotemark{g} & $7.3_{-0.9}^{+1.0}$ & $1.01_{-0.11}^{+0.17}$\\  
J1306$+$0356C & 13:06:08.3271(14) & $+$03:56:26.128(21) & $6.03519_{-0.00012}^{+0.00013}$ & ---\tablenotemark{g} & $72_{-6}^{+6}$ & $<$$38$ & $206_{-15}^{+16}$ & $5.8_{-0.9}^{+1.0}$ & $1.07_{-0.12}^{+0.14}$\\ 
\hline
PJ231$-$20Q & 15:26:37.8403(2) & $-$20:50:00.790(3) & $6.58734_{-0.00008}^{+0.00008}$ & $<$$120$ & $150_{-3}^{+3}$ & $33_{-7}^{+5}$ & $111_{-12}^{+9}$ & $11.3_{-0.7}^{+0.7}$ & $0.54_{-0.03}^{+0.03}$\\  
PJ231$-$20C & 15:26:37.8721(5) & $-$20:50:02.425(6) & $6.59074_{-0.00015}^{+0.00015}$ & $>$$210$ & $203_{-7}^{+7}$ & $<$$22$ & $250_{-9}^{+9}$ & $3.57_{-0.21}^{+0.22}$ & $0.82_{-0.04}^{+0.04}$\\  
\hline
J2100$-$1715Q & 21:00:54.6996(11) & $-$17:15:22.008(15) & $6.08142_{-0.00025}^{+0.00024}$ & ---\tablenotemark{g} & $147_{-10}^{+11}$ & $<$$40$ & $30_{-40}^{+30}$ & $4.3_{-0.5}^{+0.6}$ & $0.90_{-0.09}^{+0.10}$\\  
J2100$-$1715C & 21:00:55.4197(8) & $-$17:-15:-22.124(19) & $6.0806_{-0.0004}^{+0.0004}$ & $206_{-28}^{+26}$ & $189_{-18}^{+19}$ & $67_{-5}^{+4}$ & $13_{-4}^{+4}$ & $5.5_{-0.9}^{+1.1}$ & $1.15_{-0.12}^{+0.13}$\\  
\cutinhead{Dispersion-dominated model}
J0842$+$1218Q & 08:42:29.4380(7) & $+$12:18:50.434(12) & $6.07638_{-0.00021}^{+0.00021}$ & & $163_{-8}^{+10}$ & & & $4.2_{-0.6}^{+0.7}$ & $0.88_{-0.09}^{+0.10}$\\  
J0842$+$1218C & 08:42:28.9743(7) & $+$12:18:54.965(11) & $6.06664_{-0.00017}^{+0.00018}$ & & $148_{-8}^{+8}$ & & & $5.2_{-0.6}^{+0.7}$ & $0.82_{-0.06}^{+0.07}$\\ 
\hline
PJ167$-$13Q & 11:10:33.9822(6) & $-$13:29:45.865(7) & $6.51635_{-0.00011}^{+0.00010}$ & & $184_{-4}^{+4}$ & & & $4.7_{-0.3}^{+0.2}$ & $1.43_{-0.05}^{+0.07}$\\  
PJ167$-$13C & 11:10:34.031(2) & $-$13:29:46.27(2) & $6.5123_{-0.0002}^{+0.0002}$ & & $185_{-11}^{+14}$ & & & $2.9_{-0.3}^{+0.4}$ & $1.24_{-0.14}^{+0.17}$\\ 
\hline
J1306$+$0356Q & 13:06:08.2645(10) & $+$03:56:26.227(14) & $6.03383_{-0.00013}^{+0.00014}$ & & $107_{-6}^{+7}$ & & & $8.0_{-1.0}^{+1.0}$ & $1.03_{-0.09}^{+0.10}$\\  
J1306$+$0356C & 13:06:08.3268(15) & $+$03:56:26.127(21) & $6.03517_{-0.00013}^{+0.00013}$ & & $76_{-5}^{+6}$ & & & $5.2_{-0.8}^{+1.0}$ & $1.27_{-0.15}^{+0.17}$\\  
\hline
PJ231$-$20Q & 15:26:37.8404(2) & $-$20:50:00.789(3) & $6.58736_{-0.00007}^{+0.00007}$ & & $148_{-3}^{+3}$ & & & $13.9_{-0.8}^{+0.8}$ & $0.519_{-0.017}^{+0.018}$\\  
PJ231$-$20C & 15:26:37.8724(4) & $-$20:50:02.423(6) & $6.59065_{-0.00015}^{+0.00016}$ & & $212_{-7}^{+8}$ & & & $3.71_{-0.24}^{+0.27}$ & $0.90_{-0.05}^{+0.05}$\\
\hline
J2100$-$1715Q & 21:00:54.6990(11) & $-$17:15:22.011(13) & $6.08143_{-0.00023}^{+0.00022}$ & & $150_{-10}^{+11}$ & & & $4.5_{-0.6}^{+0.7}$ & $0.99_{-0.09}^{+0.10}$\\  
J2100$-$1715C & 21:00:55.4199(9) & $-$17:15:22.105(20) & $6.0809_{-0.0004}^{+0.0004}$ & & $245_{-17}^{+21}$ & & & $4.0_{-0.5}^{+0.6}$ & $0.88_{-0.08}^{+0.08}$\\ 
\enddata
\tablecomments{$^{\rm a}$Circular velocity. $^{\rm b}$Velocity dispersion. $^{\rm c}$Inclination. $^{\rm d}$Position angle. $^{\rm e}$Central flux density. $^{\rm f}$Exponential scale length. $^{\rm g}$ These parameters are not well determined because either the galaxy is viewed face-on and/or their is not enough evidence for rotation. For these systems the dispersion model provides a similar fit to the data.}
\end{deluxetable*}
\end{rotatetable*}

\appendix

\section{Enlarged Continuum and Moment Maps}
\label{sec:ApxMomMap}
\renewcommand\thefigure{\thesection.\arabic{figure}}  
\setcounter{figure}{0}  
\begin{figure*}[b]
\includegraphics[width=1.02\textwidth]{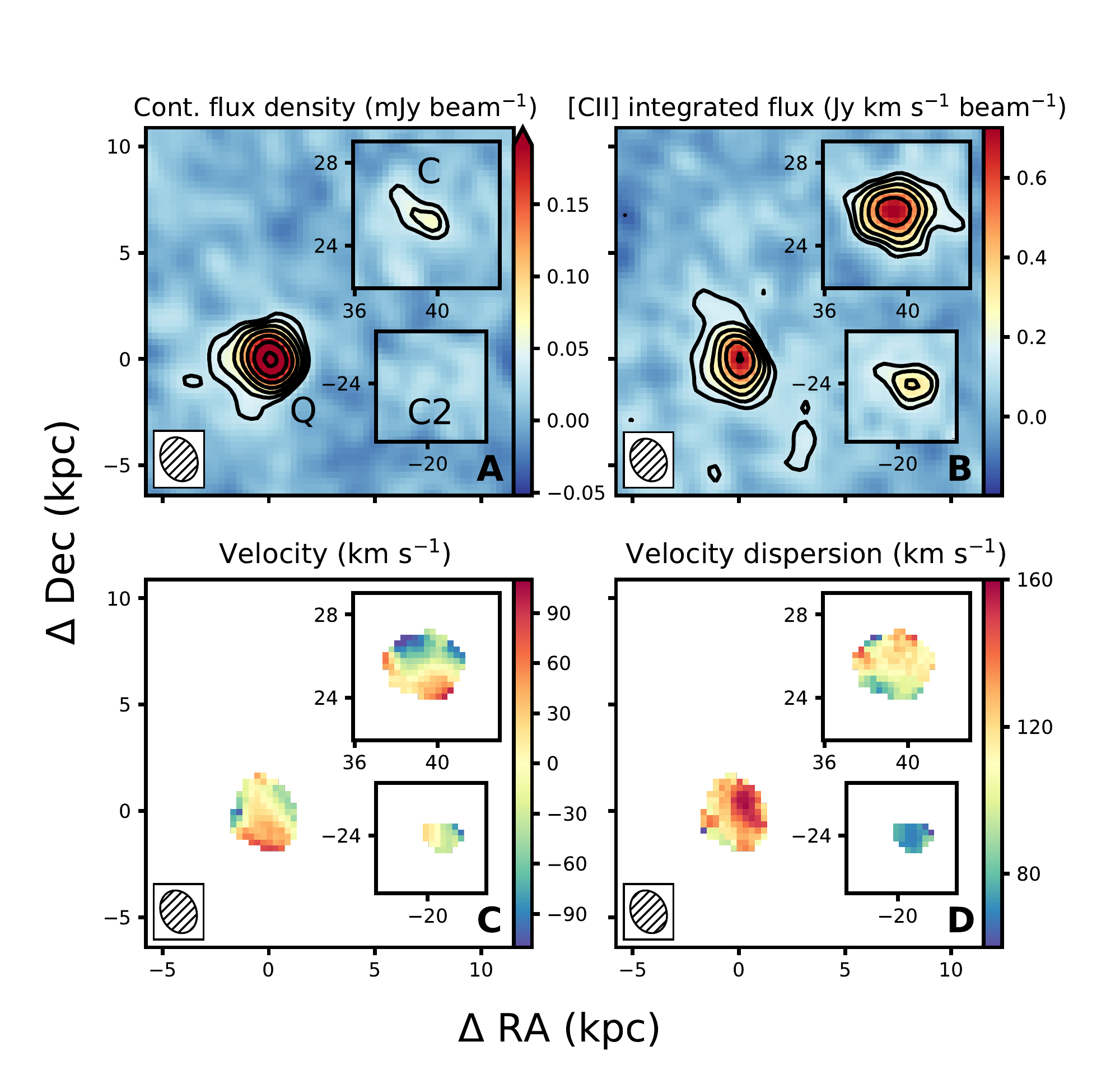}
\caption{Continuum and moment images of QSO J0842$+$1218 and its primary companion galaxy. The primary companion galaxy (shown in the top inset) is offset by 8$\farcs$2 (47~kpc)  northeast of the quasar. The second companion galaxy (shown in the bottom inset) is offset by 5$\farcs$4 (31~kpc) southwest of the quasar. Top left: 260~GHz continuum emission of the quasar and companion galaxy. Outer contours start at $3\sigma_{\rm cont}$ ($\sigma_{\rm cont}$ = 16~$\mu$Jy~beam$^{-1}$), and increase by powers of$\sqrt{2}$ for each consecutive contour. Top right: velocity-integrated flux density of the continuum-subtracted \CII\ line. Contours are defined as in the continuum panel with $\sigma_{\rm [CII]}$ = 0.040~Jy~km~s$^{-1}$~beam$^{-1}$. The bottom row shows the mean velocity field (bottom left) and the velocity dispersion (bottom right) of the \CII\ line. The mean velocity is with respect to the redshift of the quasar ($z=6.0760$) for the main figure, and with respect to the redshift of the companions (C; $z=6.0656$ and C2; $z=6.0649$) for the insets. The size of the ALMA synthesized beam is shown in the bottom left corner of each panel.}
\label{fig:J0842+12_All}
\end{figure*}
\clearpage 

\begin{figure*}[!h]
\includegraphics[width=1.02\textwidth]{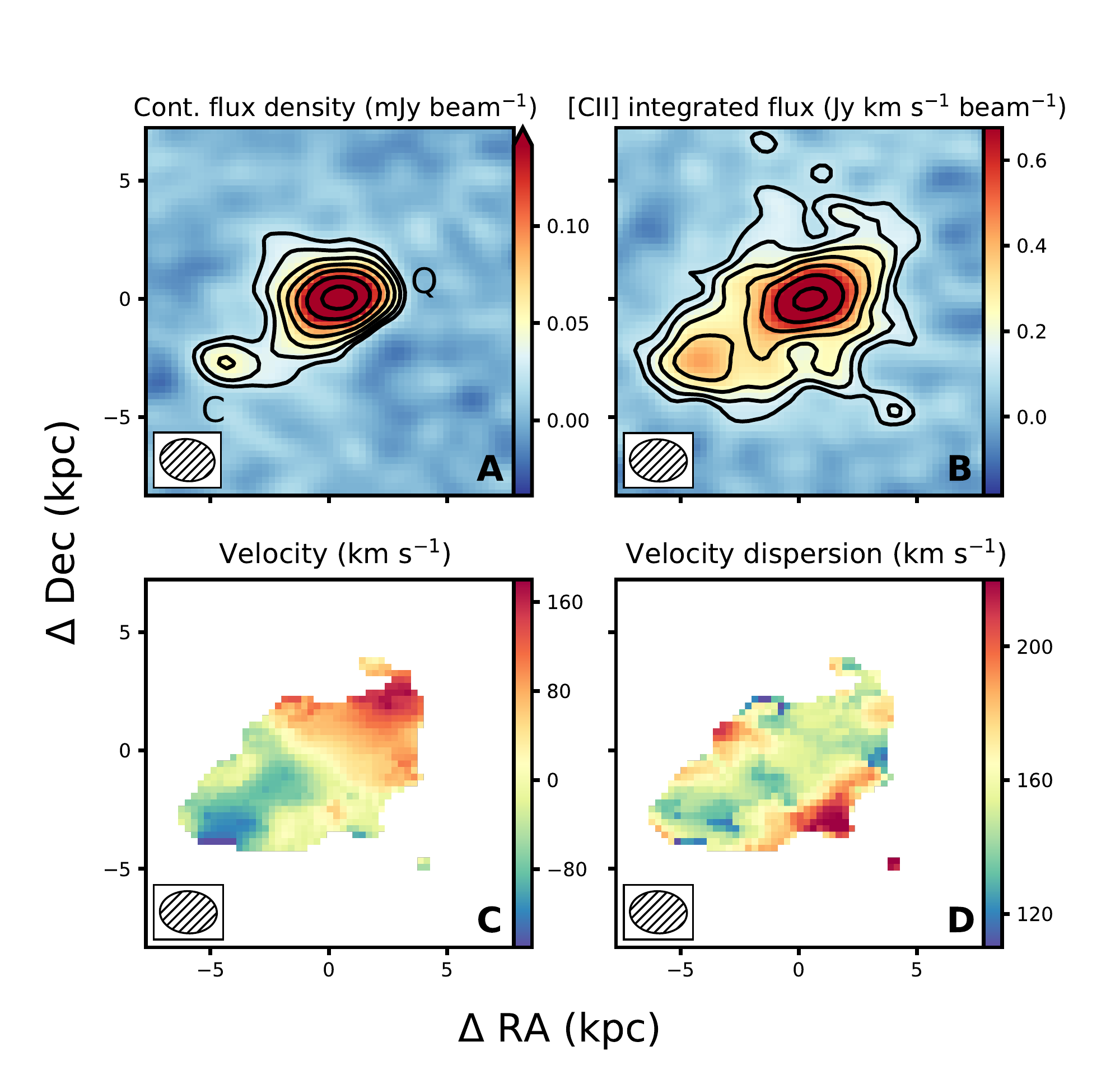}
\caption{246~GHz continuum and moment images of the quasar host and companion galaxy toward QSO PJ167$-$13. Panels and annotations are the same as those of Fig \ref{fig:J0842+12_All}, with $\sigma_{\rm cont}$ = 12~$\mu$Jy~beam$^{-1}$ and $\sigma_{\rm [CII]}$ = 0.038~Jy~km~s$^{-1}$~beam$^{-1}$. The mean velocity is with respect to the redshift of the quasar ($z = 6.5154$).
\label{fig:PJ167-13_All}}
\end{figure*}
\clearpage 

\clearpage 
\begin{figure*}[!h]
\includegraphics[width=1.02\textwidth]{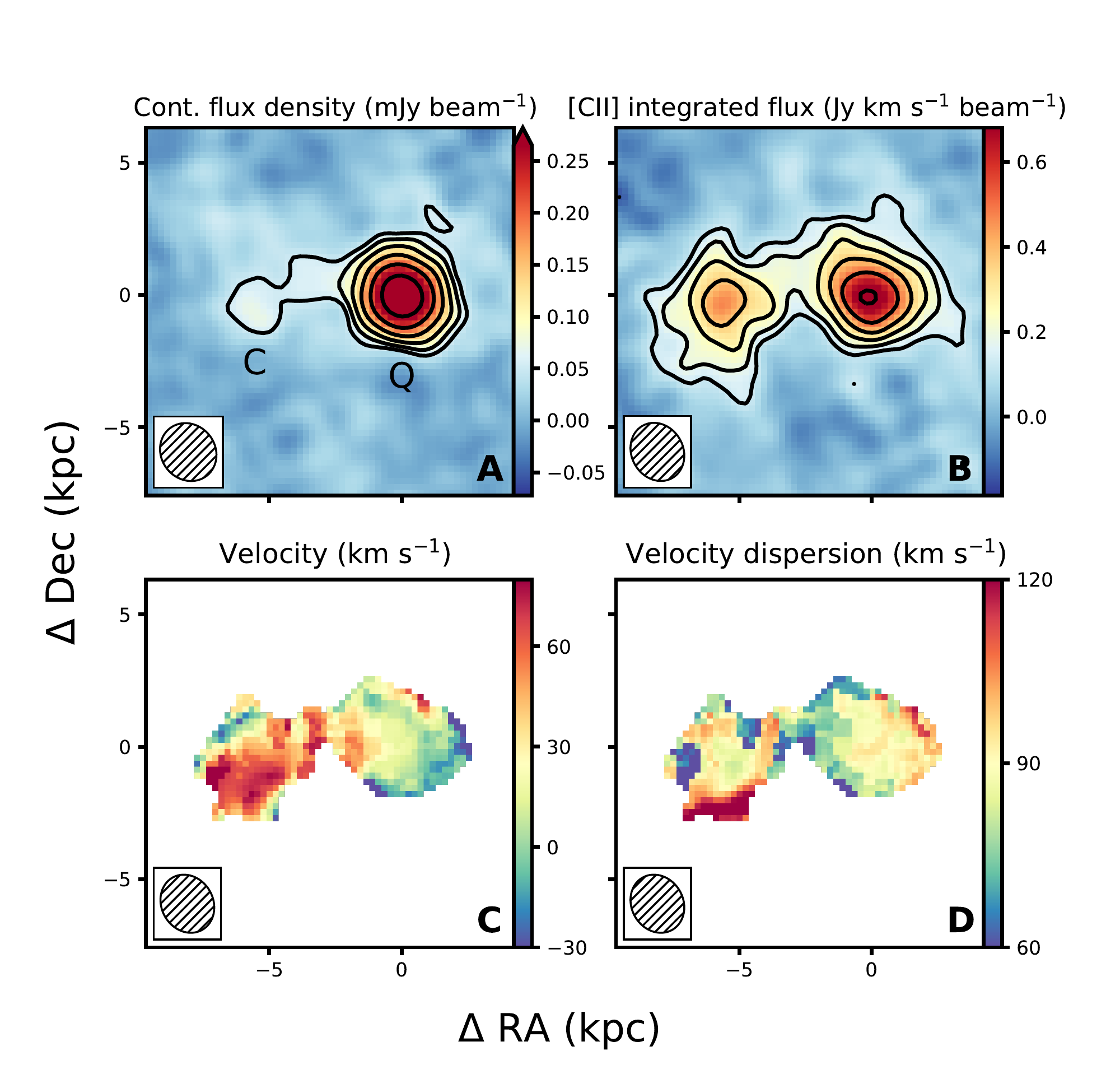}
\caption{263~GHz continuum and moment images of the quasar host and companion galaxy toward QSO J1306$+$0356. Panels and annotations are the same as those of Fig \ref{fig:J0842+12_All}, with $\sigma_{\rm cont}$ = 23~$\mu$Jy~beam$^{-1}$ and $\sigma_{\rm [CII]}$ = 0.044~Jy~km~s$^{-1}$~beam$^{-1}$. The mean velocity is with respect to the redshift of the quasar ($z = 6.0328$).
\label{fig:J1306+0356_All}}
\end{figure*}
\clearpage

\begin{figure*}[!h]
\includegraphics[width=1.02\textwidth]{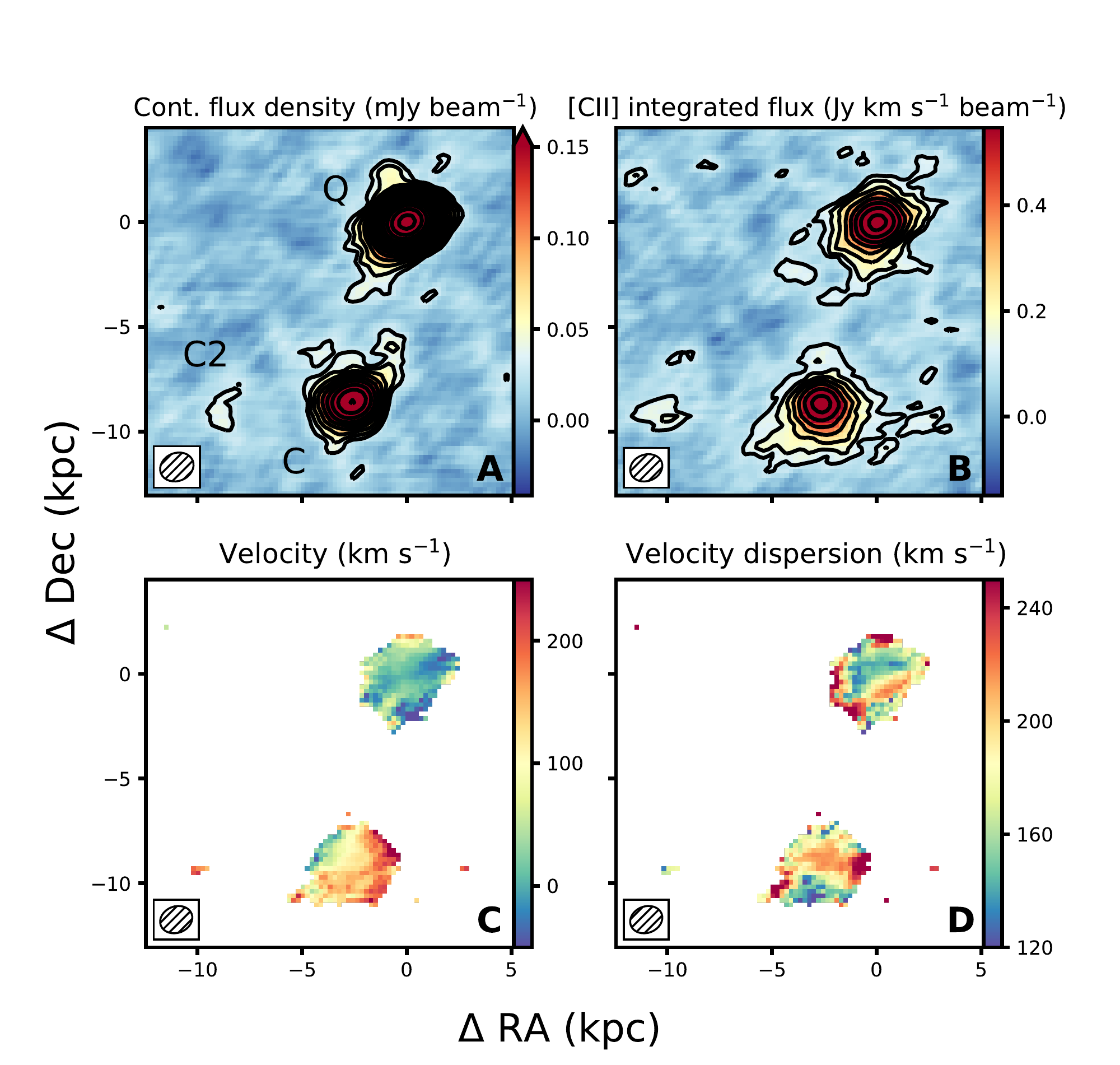}
\caption{243~GHz continuum and moment images of the quasar host and companion galaxies toward QSO PJ231$-$20. Panels and annotations are the same as those of Fig \ref{fig:J0842+12_All}, with $\sigma_{\rm cont}$ = 12~$\mu$Jy~beam$^{-1}$ and $\sigma_{\rm [CII]}$ = 0.036~Jy~km~s$^{-1}$~beam$^{-1}$. The mean velocity is with respect to the redshift of the quasar ($z = 6.5867$). The second companion, C2, can be seen in both the continuum and the \CII\ maps and is offset due east from the primary companion galaxy by $1\farcs1$ (6~kpc).
\label{fig:PJ231-20_All}}
\end{figure*}
\clearpage 

\begin{figure*}[!h]
\includegraphics[width=1.02\textwidth]{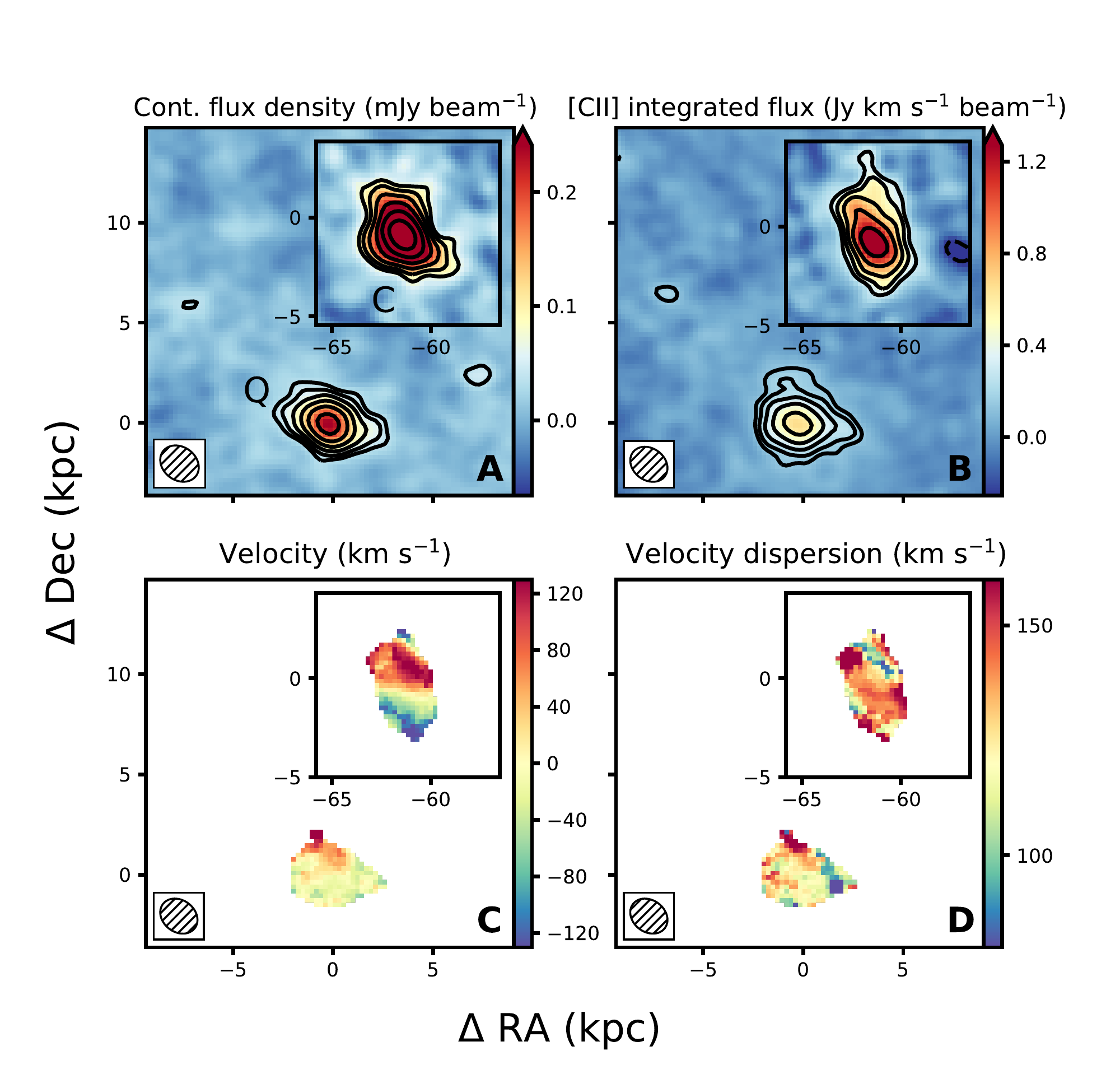}
\caption{260~GHz continuum and moment images of the quasar host and companion galaxy toward quasar QSO J2100$-$1715. The companion galaxy is offset from the quasar host by $10\farcs8$ (61~kpc) and is shown in the inset. Panels and annotations are the same as those of Fig \ref{fig:J0842+12_All}, with $\sigma_{\rm cont}$ = 21~$\mu$Jy~beam$^{-1}$ and $\sigma_{\rm [CII]}$ = 0.047~Jy~km~s$^{-1}$~beam$^{-1}$. The mean velocity is with respect to the redshift of the quasar ($z=6.0809$) for the main figure, and with respect to the redshift of the companion ($z=6.0814$) for the inset.
\label{fig:J2100-1715_All}}
\end{figure*}
\clearpage 

\section{Channel Maps}
\label{sec:ApxChanMap}
\setcounter{figure}{0}  
\begin{figure*}[!h]
\includegraphics[width=1.02\textwidth]{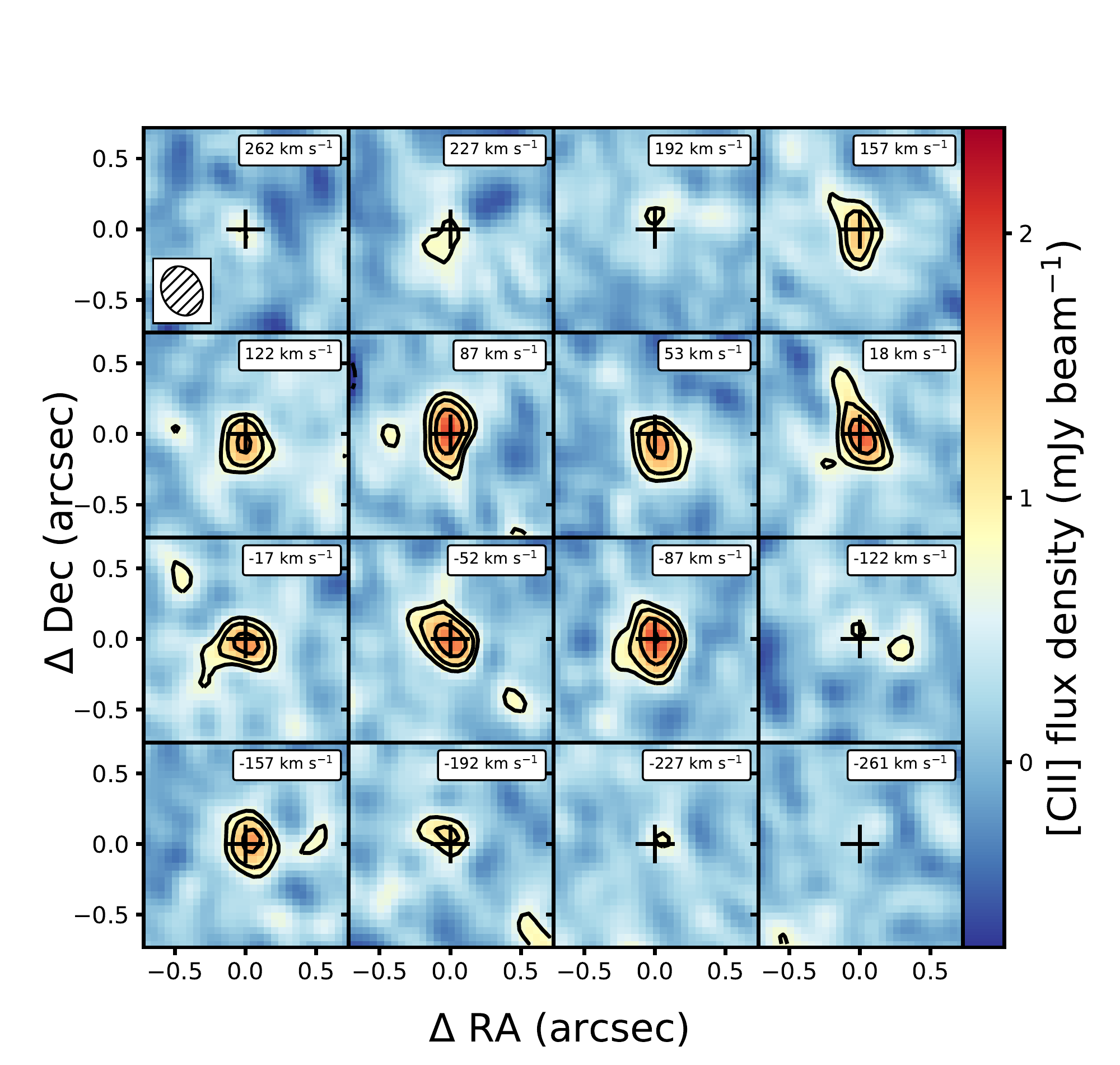}
\caption{Channel maps of the \CII\ line for the host galaxy of quasar QSO~J0842$+$1218. Velocities of each channel map are relative to the central redshift of the \CII\ emission as determined from a Gaussian fit to the data ($z_{\rm [CII]}~=~6.0760$). The ALMA synthesized beam is shown in the inset of the top left plot. The plus-sign marks the position of the continuum emission from this source as determined from a 2D Gaussian fit to the data (Table \ref{tab:obsfirprop}). Contours start at 3$\sigma$ and consecutive contours increase by powers of $\sqrt{2}$, where $\sigma = 0.23$~mJy~beam$^{-1}$.
\label{fig:J0842+12QSO_CM}}
\end{figure*}
\clearpage

\begin{figure*}[!h]
\includegraphics[width=1.02\textwidth]{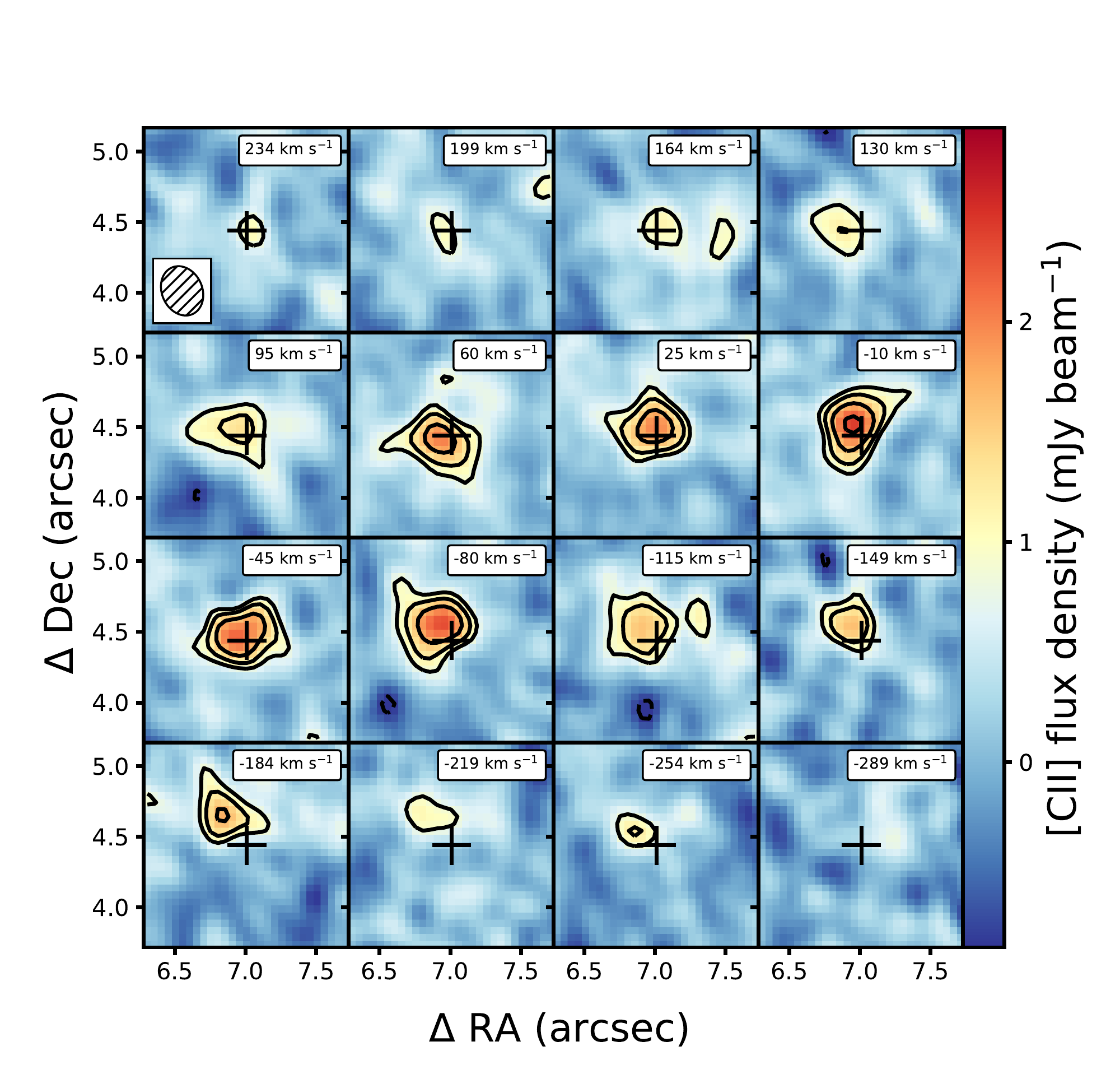}
\caption{Channel maps of the \CII\ line for the primary companion galaxy toward quasar QSO~J0842$+$12. Velocities of each channel map are relative to the redshift of the \CII\ emission from the galaxy ($z_{\rm [CII]}~=~6.0656$). Annotations and contours are the same as Figure \ref{fig:J0842+12QSO_CM}, with $\sigma = 0.23$~mJy~beam$^{-1}$.
\label{fig:J0842+12COMP_CM}}
\end{figure*}
\clearpage

\begin{figure*}[!h]
\includegraphics[width=1.02\textwidth]{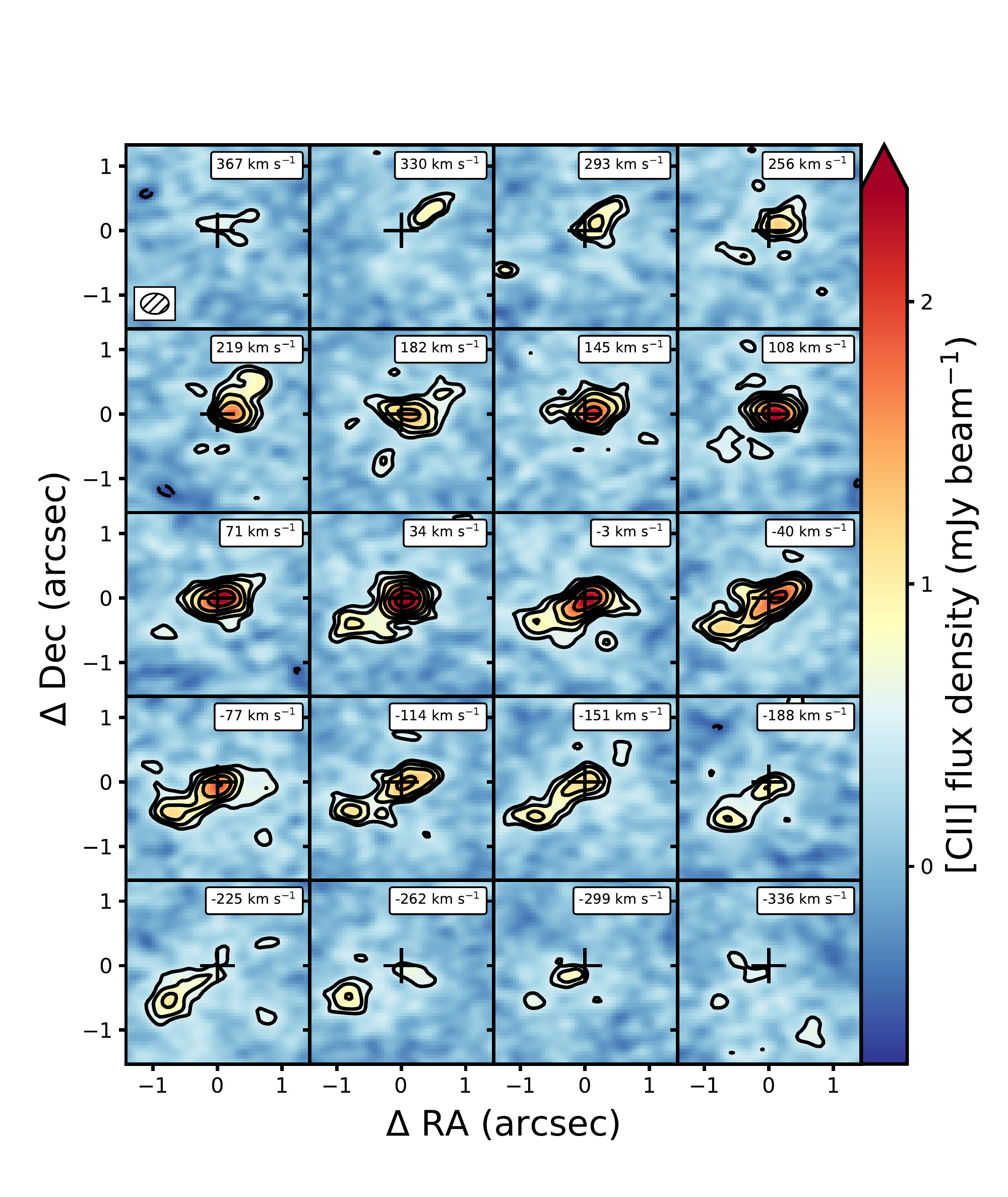}
\caption{Channel maps of the \CII\ line for the quasar host and companion galaxy of QSO~PJ167$-$13. Velocities of each channel map are relative to the central redshift of the \CII\ emission from the quasar host ($z_{\rm [CII]}~=~6.5154$). Annotations and contours are the same as Figure \ref{fig:J0842+12QSO_CM}, with $\sigma = 0.16$~mJy~beam$^{-1}$.
\label{fig:PJ167-13_CM}}
\end{figure*}
\clearpage

\begin{figure*}[!h]
\includegraphics[width=1.02\textwidth]{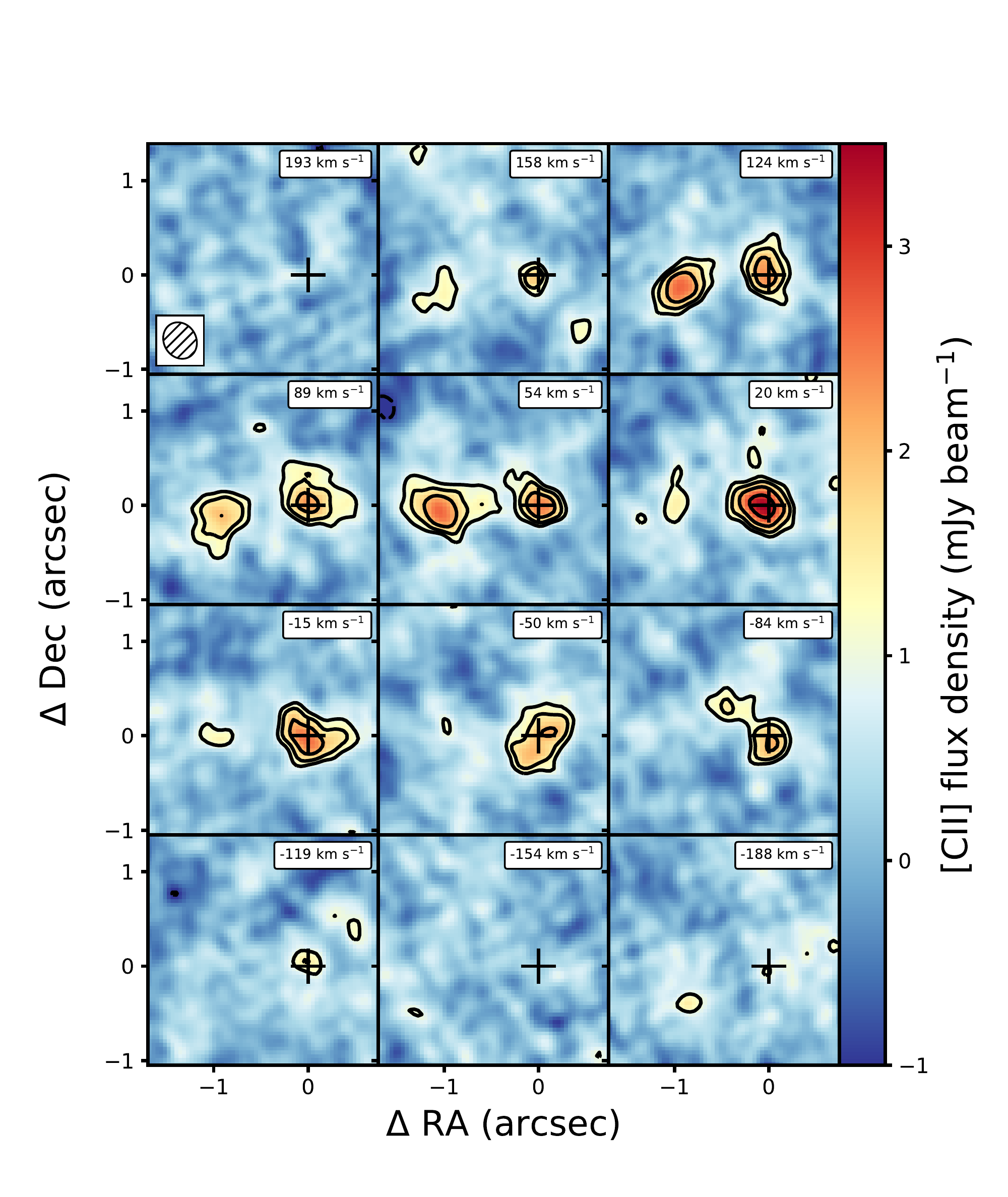}
\caption{Channel maps of the \CII\ line for the quasar host and companion galaxy of QSO~J1306$+$0356. Velocities of each channel map are relative to the central redshift of the \CII\ emission from the quasar host ($z_{\rm [CII]}~=~6.0328$). Annotations and contours are the same as Figure \ref{fig:J0842+12QSO_CM}, with $\sigma = 0.33$~mJy~beam$^{-1}$.
\label{fig:J1306+0356_CM}}
\end{figure*}
\clearpage

\begin{figure*}[!h]
\includegraphics[width=1.02\textwidth]{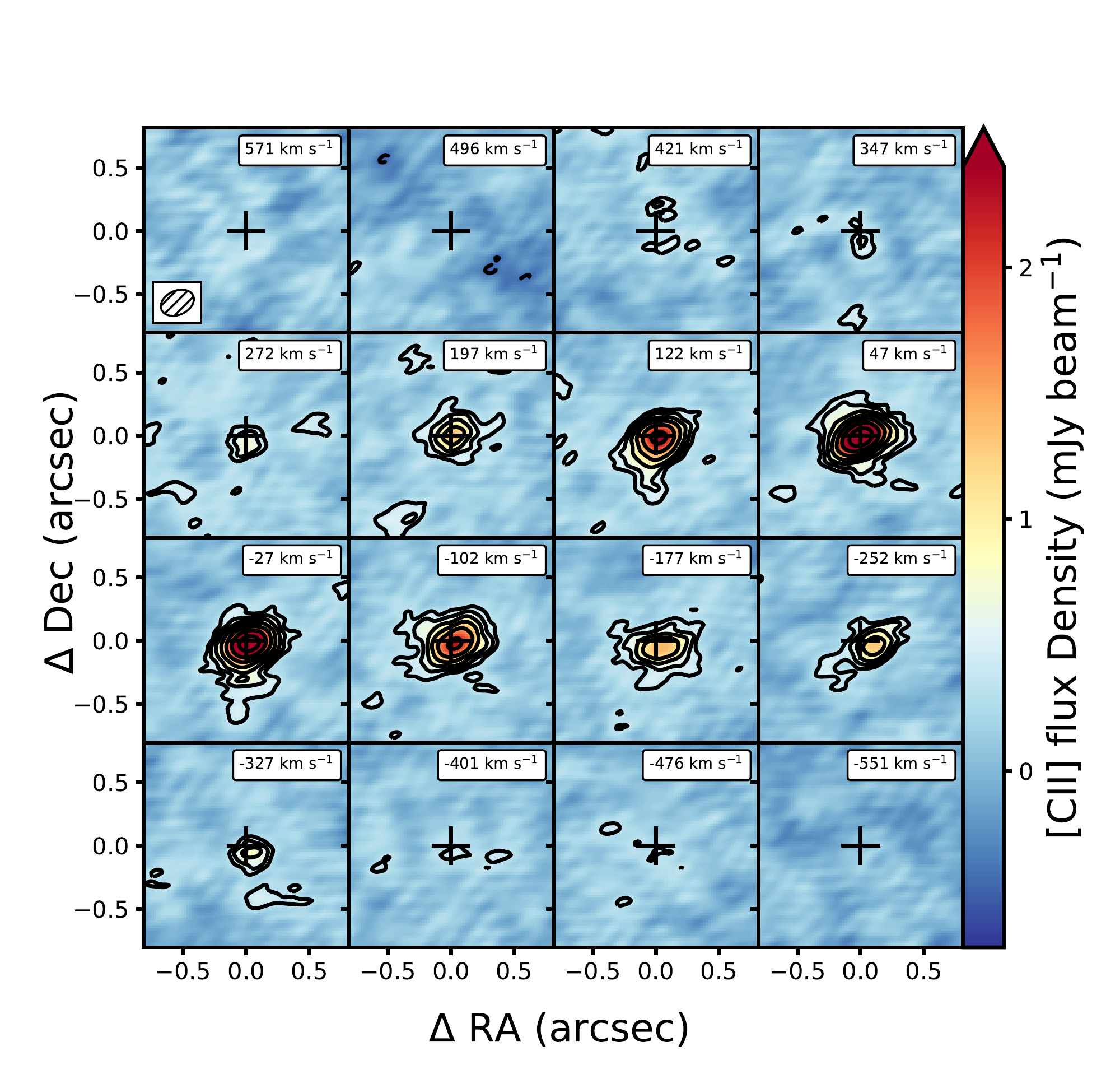}
\caption{Channel maps of the \CII\ line for the quasar host galaxy of QSO~PJ231$-$20. Velocities of each channel map are relative to the central redshift of the \CII\ emission from the quasar host galaxy ($z_{\rm [CII]}~=~6.5867$). Annotations are the same as Figure \ref{fig:J0842+12QSO_CM}, with $\sigma = 0.12$~mJy~beam$^{-1}$. Because of the width of the \CII\ emission, we have resampled the data by 2 channels ($\approx$62~MHz or $\approx$75~\kms) for visual purposes.
\label{fig:PJ231-20QSO_CM}}
\end{figure*}
\clearpage

\begin{figure*}[!h]
\includegraphics[width=1.02\textwidth]{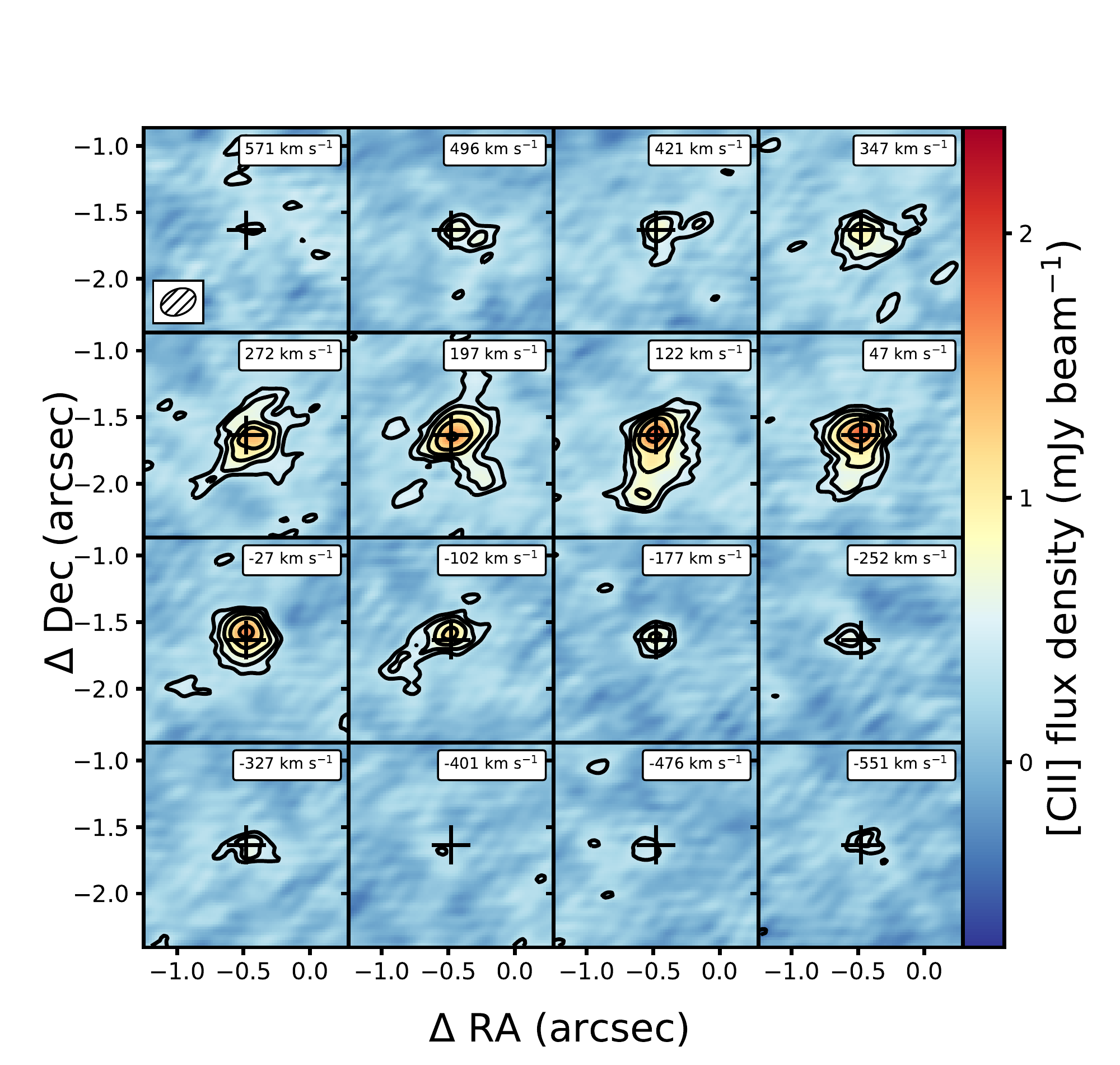}
\caption{Channel maps of the \CII\ line for the companion galaxy of QSO~PJ231$-$20. Velocities of each channel map are relative to the central redshift of the \CII\ emission from the companion galaxy ($z_{\rm [CII]}~=~6.5901$). Annotations are the same as Figure \ref{fig:J0842+12QSO_CM}, with $\sigma = 0.12$~mJy~beam$^{-1}$. Because of the width of the \CII\ emission, we have resampled the data by 2 channels ($\approx$62~MHz or $\approx$75~\kms) for visual purposes.
\label{fig:PJ231-20COMP_CM}}
\end{figure*}
\clearpage

\begin{figure*}[!h]
\includegraphics[width=1.02\textwidth]{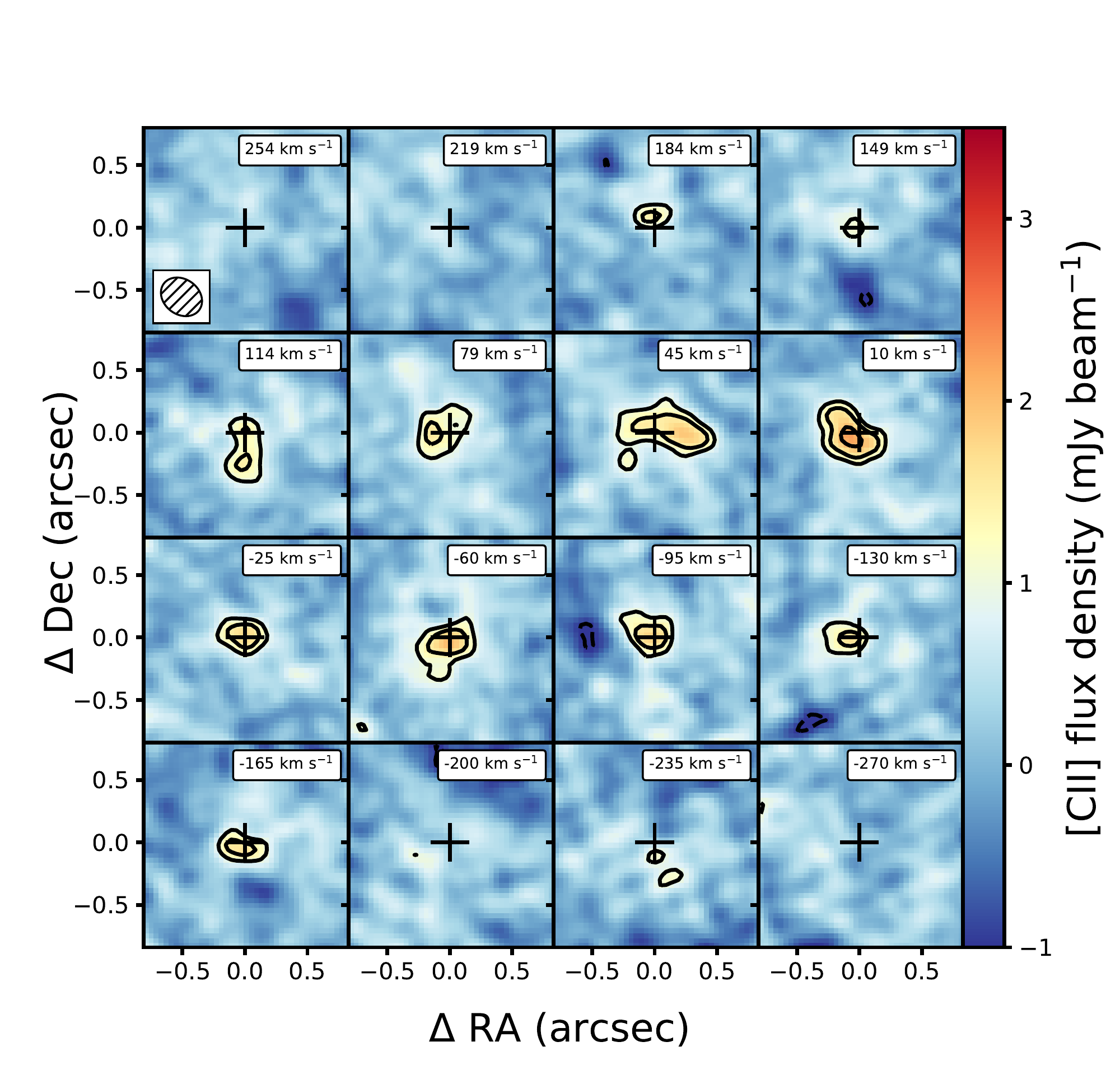}
\caption{Channel maps of the \CII\ line for the quasar host galaxy of QSO~J2100$-$1715. Velocities of each channel map are relative to the central redshift of the \CII\ emission from the quasar host galaxy ($z_{\rm [CII]}~=~6.0809$). Annotations are the same as Figure \ref{fig:J0842+12QSO_CM}, with $\sigma = 0.33$~mJy~beam$^{-1}$.
\label{fig:J2100-1715QSO_CM}}
\end{figure*}
\clearpage

\begin{figure*}[!h]
\includegraphics[width=1.02\textwidth]{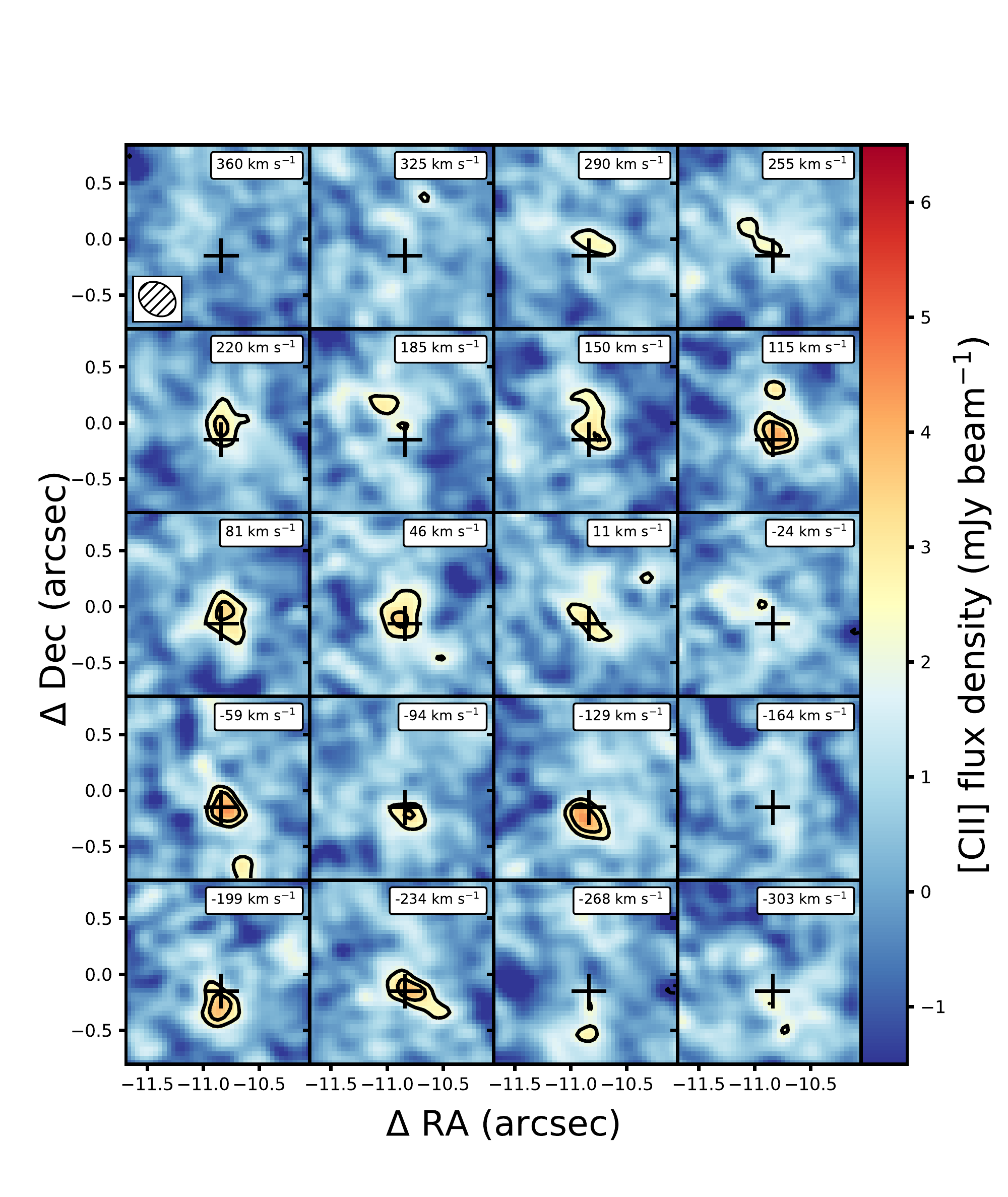}
\caption{Channel maps of the \CII\ line for the companion galaxy of QSO~J2100$-$1715. Velocities of each channel map are relative to the central redshift of the \CII\ emission of the companion galaxy ($z_{\rm [CII]}~=~6.0814$). Annotations are the same as Figure \ref{fig:J0842+12QSO_CM}, with $\sigma = 0.33$~mJy~beam$^{-1}$.
\label{fig:J2100-1715COMP_CM}}
\end{figure*}
\clearpage

\section{Models Used for Kinematic Analysis}
\label{sec:ApxModel}

To fit the kinematics of the \CII\ line, we model the line with two different models, a thin-disk model and a dispersion-dominated model. For both models we assume that the \CII\ flux density can be approximated by an exponential profile of the form:
\begin{equation}
I(R) = I_0 e^{R/R_d},
\label{eq:sup1}
\end{equation}
where $R$ is the galactocentric radius, $I_0$ is the maximum flux density at the center of the source, and $R_d$ is the scale length of the exponential function. We also assume in both models that the velocity dependence is Gaussian with constant velocity dispersion $\sigma_v$:
\begin{equation}
I(R, v) = I(R) e^{(v - v_{\rm obs})^2 / 2\sigma_v^2}
\end{equation}

For the thin-disk model, we further assume the velocities are all in the plane of disk, which has an inclination, $i$, and position angle, $\alpha$, with respect to the plane of the sky. This case has been discussed in detail in \citet{Neeleman2016}. To be specific, the galactocentric radius, $R$, is related to the projected distance, $\rho$, by \citep[see e.g.,][]{Chen2005}:
\begin{equation}
R = \rho \times \sqrt{1 + \sin^2(\phi - \alpha) \tan^2(i)}.
\end{equation}
where, $\rho$ is measured with respect to the kinematic center of the observations:  $\rho = \sqrt{(x - x_c)^2 + (y - y_c)^2}$. In this model, the observed projected velocities, $v_{\rm obs}$, are related to the rotational velocity, $v_{\rm circ}$ by:
\begin{equation}
v_{\rm obs} = \frac{\cos(\phi-\alpha)\sin(i)}{\sqrt{1 + \sin^2(\phi - \alpha)\tan^2(i)}}v_{\rm circ} + v_c,
\end{equation}
where $v_c$ is the velocity offset of the kinematic center, given as a redshift, $z_{\rm kin}$ in Table \ref{tab:kinfit}. These 9 parameters ($x_c, y_c, z_{\rm kin}, v_{\rm circ}, \sigma_v, i, \alpha, I_0, R_{\rm d}$) thus uniquely determine the flux at each position ($x, y, v$) in the model cube. 

For the dispersion-dominated model, we generate a model data cube\footnote{This is a four dimensional matrix or array, which sometimes is referred to as a `hypercube'} similar to the data cube, but with a third spatial dimension, $z$. This $z$ is chosen to be large enough, both in the positive and negative direction, to make sure the flux is not substantially truncated in the z-direction. Practically this is often similar in size as the other two dimensions. The galactocentric radius, $R$, in this case is simply:
\begin{equation}
R = \sqrt{(x - x_c)^2 + (y - y_c)^2 + z^2}.
\end{equation}
The observed velocity, $v_{\rm obs}$, is set by-definition to the systemic velocity, $v_c$, of the kinematic center of the \CII\ line in this model. Fluxes are measured at each position using equation \ref{eq:sup1}, and the cube is then collapsed (i.e., the fluxes are summed) along the z-axis to produce the final model cube (in $x, y, v$). This model is dependent on 6 parameters ($x_c, y_c, z_{\rm kin}, \sigma_v, I_0, R_{\rm d}$).

The model cubes from these two models are then used in the Markov Chain Monte Carlo method to determine the best-fit and uncertainties for the parameters of the model. The results are tabulated in Table \ref{tab:kinfit}.

\end{document}